\documentclass[%
 aip,
 amsmath,amssymb,
 reprint,%
]{revtex4-1}

\usepackage{graphicx}% Include figure files

\usepackage{subcaption}% PUESTO POR MI

\usepackage{dcolumn}% Align table columns on decimal point
\usepackage{bm}% bold math
\usepackage[utf8]{inputenc}
\DeclareUnicodeCharacter{2009}{\,}
\usepackage[T1]{fontenc}
\usepackage{mathptmx}
\usepackage{etoolbox}
\usepackage{color}

\makeatletter
\def\@email#1#2{%
 \endgroup
 \patchcmd{\titleblock@produce}
  {\frontmatter@RRAPformat}
  {\frontmatter@RRAPformat{\produce@RRAP{*#1\href{mailto:#2}{#2}}}\frontmatter@RRAPformat}
  {}{}
}%
\makeatother
\begin{document}

\title
{Mobility-induced phase separation in a binary mixture of active Brownian particles}
% Force line breaks with \\
\author{D. Jim\'enez-Flores}
\affiliation{Departamento de F\'{\i}sica At\'omica, Molecular y Nuclear, \'Area de F\'{\i}sica Te\'orica, Facultad de F\'{\i}sica, Universidad de Sevilla, Avenida de Reina Mercedes s/n, 41012 Sevilla (Spain)}
\author{A. Rodr\'{\i}guez-Rivas}%
 %\email{Second.Author@institution.edu.}
\affiliation{ 
Departamento de Matem\'atica Aplicada II, Universidad de Sevilla, E.T.S. de Ingenier\'{\i}a, C. de los Descubrimientos, s/n. Pabell\'on Pza. de Am\'erica
41092 Sevilla (Spain) %\\This line break forced with \textbackslash\textbackslash
}%

\author{J. M. Romero-Enrique}
 %\homepage{http://www.Second.institution.edu/~Charlie.Author.}
\email{enrome@us.es.}
\affiliation{%
Departamento de F\'{\i}sica At\'omica, Molecular y Nuclear, \'Area de F\'{\i}sica Te\'orica, Facultad de F\'{\i}sica, Universidad de Sevilla, Avenida de Reina Mercedes s/n, 41012 Sevilla (Spain)}
\affiliation{
Instituto Carlos I de F\'{\i}sica Te\'orica y Computacional, Campus Universitario Fuentenueva
Calle Dr. Severo Ochoa, 18071 Granada (Spain)
%\\This line break forced% with \\
}%

\date{\today}% It is always \today, today,

\begin{abstract}

In this paper, we report a Brownian dynamics simulation of the mobility-induced phase separation which occurs in a two-dimensional binary mixture of active soft Brownian particles, whose interactions are modeled by non-additive Weeks-Chandler-Andersen potentials inspired in Lennard-Jones potentials used for glass-forming passive mixtures. The analysis of structural properties, such as the radial distribution functions and the hexatic order parameter, shows that the high-density coexisting state in the binary case is spatially disordered, unlike the solid-like state observed for the monocomponent system. Characterization of the mean-square displacement of the active particles shows that both the low- and high-density coexisting states have diffusive behavior for long times. Thus, the high-density coexisting states are liquid-like in the binary cases. Moreover, diffusive behavior is also observed in the high-density solid-like state for the monocomponent system, which is driven by the presence of active topological defects.

\end{abstract}

\maketitle

\section{\label{introduction} Introduction}

The study of active matter constitutes a rapidly growing field in physics, being an essential generic model for systems that are intrinsically out-of-equilibrium \cite{MJRLPRS13,R17,DSM20}. Composed of self-propelled agents, this matter presents many examples in nature, ranging from biological tissues \cite{PABGKBCT19}, biofilms \cite{S20}, bird flocks\cite{Ballerini2008}, fish schools\cite{Tunstrom2013} animal herds\cite{GUERON199685}, and pedestrian crowds \cite{GGBZB25}, and in synthetic systems, like phoretic Janus particles\cite{Paxton2006,Howse2007}, small robots\cite{Li_2013,Li2019}, and vibrated granular matter\cite{Narayan2007,Deseigne2010}, exhibiting complex emergent collective behavior.

From a theoretical point of view, different agent-based models have been proposed since the seminal work by Vicsek and coworkers \cite{Vicsek1995}. Focusing on scalar active matter, in which the only hydrodynamic mode is the conserved density field \cite{Granek_Colloquium_2024}, the standard models are the Active Brownian Particle (ABP) model \cite{Fily2012,RBELS12}, the Run-and-Tumble Particle (RTP) model \cite{Schnitzer1993,SCT15,SBS20} and the Active Ornstein-Uhlenbeck Particle model\cite{Hanggi1994}. In both ABP and RTP particles have a self-propelling force of constant norm and random orientation. In the ABP model, particle orientations change continuously due to rotational diffusion. In the RTP model, by contrast, particles reorient instantaneously at random instants or ``rumbles'' occurring at constant rate. Finally, in the AOUP model both norm and orientation fluctuate following an Ornstein-Uhlenbeck process. The presence of interactions between the particles, either by conservative pairwise forces or more complex mechanisms such as the quorum sensing\cite{Miller2001}, are fundamental to understand the collective behavior of the scalar active matter. %. 
Scalar active matter may present phase transitions such as the motility-induced phase separation (MIPS). In this transition, two flux-free, homogeneous and stationary states characterized by different densities coexist\cite{Tailleur_RunTumble_2008,Fily2012,Redner2013,CT15}. The existence of this transition, originally observed in computer simulations, was also supported by experiments\cite{Buttinoni2013}, Cahn-Hilliard-like phase-field models \cite{Wittkowski2014,Speck2014, CT15,Solon2018,Paliwal_ChemPot_2018} and mechanical approaches \cite{Omar_MechanicalTheory_2023}. Despite that, in some respects, MIPS may resemble the well-known vapor-liquid transition in equilibrium fluids, its origin is quite different. In fact, MIPS does not require attraction between the particles, unlike its equilibrium counterpart: it results from the positive feedback between the tendency of the active particles to accumulate where they move slower and their slowdown at high densities due to collisions and repulsive forces\cite{CT15}. On the other hand, the dense stationary state usually shows a hexatic- or solid-like structure\cite{Redner2013,Digregorio2018,CARTV22}.

 Although many of the simulation studies consider monodisperse systems, in the recent years the effect of the polydispersity on the size and/or the self-mobility of the active particles has been explored\cite{CDSS21,KSGM21,KJB22,CL22,RCS23,SJBS23,PF25,BFJDI25}. For example, MIPS has been studied in isometric mixtures of active and passive Brownian particles within the ABP model\cite{Stenhammart2015,Wysocki_2016,C9SM01803D,Fernandez-Quevedo_2025}. These studies show a shift to higher self-propelling forces and densities of the MIPS coexistence curve, segregation of the active particles to the interfacial region in the dense phase and propagating interfaces due to the flux imbalance of the active and passive particles in the dilute phase. On the other hand, the effect of size polydispersity of ABP model of Yukawa particles \cite{SCD24} and pseudo-hard spheres\cite{TS24}, keeping the same self-propelling forces and diffusive parameters for all particles, has also been studied. However, these studies do not directly address its effect on the MIPS coexistence curve. In this paper, we present a preliminary study of a binary mixture of ABPs with Weeks-Chandler-Andersen-like interactions. We use a set of potential parameters for the interactions for each pair of particles based on the Lennard-Jones potentials in a glass-forming passive binary mixture \cite{KA94}, in order to check if spatial ordering in the high-density coexisting phase (observed in the dense phase in the monocomponent case) can be inhibited. We will also study the dynamical properties to discern between liquid-like and glassy-like states. 

This paper is organized as follows. In Section \ref{methodology}, we provide details on the simulation model and methodology used for this work. In Section \ref{results}, we show the data obtained from simulations and analyze the results focusing on the structural properties such as the radial distribution functions and the hexatic order parameter, as well as dynamical properties such as the mean-square displacements of the active particles. Finally, in Section \ref{conclusions}, we present the conclusions of our study.

\section{\label{methodology} Methodology}

We employ numerical Brownian dynamics simulations to study the collective behavior of mixtures of soft ABPs under conditions that favor phase separation in two-dimensional systems. To analyze the effect of composition, we consider both monocomponent and binary systems. This comparison allows us to characterize the properties of the dense and dilute phases in each case.

\subsection{Model}
Our system consists of a binary mixture of $N_1$ active particles of class 1 and $N_2$ active particles of class 2, being $N=N_1+N_2$ the total number of active particles. We describe the dynamics of each particle through the overdamped Langevin equations. Each particle $i$ is characterized by its 2D position vector $\mathbf{r}_i$ and orientation unit vector $\mathbf{e}_i=\left(\cos\theta_i,\sin\theta_i\right)$. The translational and rotational evolution equations are given by
\begin{eqnarray}
    \dot{\mathbf{r}}_i&=&v_a \mathbf{e}_i - \frac{D_S}{k_B T}\boldsymbol{\nabla}_{\mathbf{r}_i}U+ \sqrt{2D_S}\boldsymbol{\xi}_i\left(t\right),\\
    \dot{\theta_i}&=&\sqrt{2D_R}\eta_i\left(t\right),
\end{eqnarray}
where $T$ is the temperature and $v_a$ is the self-propelling speed of the particle, $U$ is the total potential energy due to interactions between the active particles. Note that the self-propelling velocity $v_a\mathbf{e}$ is related to the active force $\mathbf{F}_a$ as $\mathbf{F}_a=\gamma_S v_a\mathbf{e}$, where $\gamma_S$ is the translational friction coefficient. The translational and rotational diffusion coefficients $D_S$ and $D_R$, respectively, are related to the translational and rotational friction coefficients $\gamma_S$ and $\gamma_R$ through $D_S=k_B T/\gamma_S$ and $D_R=k_B T/\gamma_R$. The stochastic terms $\boldsymbol{\xi}_i \left(t\right)=\left(\xi_{ix}\left(t\right),\xi_{iy} \left(t\right)\right)$ and $\eta_i\left(t\right)$ represent Gaussian white noise with zero mean and correlations $\langle\xi_{i\alpha}\left(t\right) \xi_{j\beta} \left(t'\right)\rangle=\delta_{ij} \;\delta_{\alpha\beta}\;\delta\left(t-t'\right)$ y $\langle\eta_i\left(t\right) \eta_j\left(t'\right)\rangle=\delta_{ij}\;\delta\left(t-t'\right)$. Note that we assume that all particles have the same values of $v_a$, $D_S$, and $D_R$, regardless of their class. Thus, the differences between species are introduced through the interaction potential. As usual, the total potential energy is taken to be pairwise additive 
\begin{equation}
U=\frac{1}{2}\sum_{i=1}^N \sum_{j\ne i}^N V^{(\alpha_i,\beta_j)}(r_{ij}),
\label{potenergy}
\end{equation}
where $\alpha_i$ and $\beta_j$ represent the classes of the particles $i$ and $j$, respectively, $r_{ij}$ is the distance between the particles $i$ and $j$, and $V^{(\alpha,\beta)}(r)$ is the interaction potential between a particle of class $\alpha$ and a particle of class $\beta$ separated by a distance $r$. 
In this paper, we model interactions between particles through the purely repulsive Weeks–Chandler–Andersen (WCA) potential\cite{WCA}
\begin{equation}
V^{(\alpha,\beta)}(r) =
4\varepsilon_{\alpha\beta} \left[ 
\left( \dfrac{\sigma_{\alpha\beta}}{r} \right)^{12} - 
\left( \dfrac{\sigma_{\alpha\beta}}{r} \right)^{6} 
\right]+\varepsilon_{\alpha\beta}
\end{equation}
if $r<2^{1/6}\sigma_{\alpha\beta}$, and it vanishes otherwise. Here $\epsilon_{\alpha\beta}$ and $\sigma_{\alpha\beta}$ are the energy and distance scales for the interaction potential. We adopt energy and length reduced units in terms of $\epsilon_{11}\equiv \epsilon$ and $\sigma_{11}\equiv \sigma$. Thus, the temperature is expressed in units of $\epsilon/k_B$. Moreover, we choose $\epsilon_{12}=1.5\epsilon$, $\epsilon_{22}=0.5\epsilon$, $\sigma_{12}=0.8\sigma$, and $\sigma_{22}=0.88\sigma$. This choice is inspired by previous studies by Kob and Andersen\cite{KA94} that showed that crystallization is inhibited in an 80:20 binary mixture of passive Lennard-Jones particles with this parametrization. Instead, glassy behavior is observed in a wide range of temperatures.

 For the dynamic parameters, first we assume that the reduced temperature $k_B T/\epsilon=1$. We use as time unit $\tau=\sigma^2/D_S$ and suppose that the rotational and translational diffusion coefficients for particles of class 1 obey the Stokes-Einstein-like relation $D_R=3D_S/\sigma^2$. The P\'eclet number for particles of class 1, which controls their self-propulsion speed, is given by the expression $Pe=3v_a/\sigma D_R$. In reduced units, $D_S=1$, $D_R=3$ and $v_a=Pe$. As mentioned above, particles of class 2 are assumed to have the same values of $D_S$, $D_R$ and $v_a$ than particles of class 1. In this way, both classes of particles will move identically in absence of particle interactions, so we can isolate the effect of size polydispersity on the MIPS. However, this choice implies that particles of class 2 would not satisfy the Stokes-Einstein relationship and that $Pe$ is not, strictly speaking, their P\'eclet number.

\subsection{Numerical details and methodology}
We carried out Brownian dynamics simulations with reduced time step $\Delta t=10^{-6}$ by using the open-source package LAMMPS \cite{LAMMPS}. In a first stage, simulations in boxes of reduced dimensions $L_x\times L_y=240\times 40$ under periodic boundary conditions and an overall constant particle density $\rho\equiv N/L_x L_y$ were run for $1.2\times 10^9$ time steps. This anisotropic simulation box is chosen to favor the formation of dense and dilute slabs of particles under MIPS conditions, separated by interfacial regions, which corresponds to the MIPS coexisting stationary states. This procedure is analogous to the direct coexistence method for equilibrium phase transition of passive systems. The mole fractions associated with species $1$, $X_1=N_1/N$, are selected to be $1$ (pure $1$-monocomponent system), $0.75$ and $0.5$. For each value of $X_1$, a wide range of values of $Pe$ are considered. We considered overall reduced densities $\rho=0.6$ for $X_1=1$ and $X_1=0.75$, which correspond to $N_1=5760$ particles of class 1 for $X_1=1$; and $N_1=4320$ particles of class 1 and $N_2=1440$ particles of class 2. As the dense slab becomes too narrow for $X_1=0.5$ in the previous value of overall density, $\rho$ was changed to $0.8$ for this composition, i.e. $N_1=N_2=3840$ particles of each class. We start for a large value of $Pe=150$ and a configuration in which particles are placed at random positions following an uniform distribution. In order to eliminate possible overlaps between particles, we initially perform an energy minimization of the system to obtain the initial state for our simulations. Now, we run the Brownian dynamics simulation until the stationary state is reached. The occurrence of MIPS in the system is identified by monitoring total density and mole fraction profiles along the long $x-$axis
\begin{eqnarray}
\rho_\alpha(x)&=&\bigg\langle \sum_{i=1}^{N}\delta_{\alpha_i,\alpha} \delta(x-x_i+x_{CM})\bigg\rangle\ , \label{rho1profile}\\ X_1(x)&=&\frac{\rho_1(x)}{\rho_1(x)+\rho_2(x)}\, \label{X1profile}
\end{eqnarray}
where $\alpha_i$ is the class of the particle $i$, which is located at a horizontal position $x_i$, and $x_{CM}$ is the $x-$coordinate of the center of mass of the system. A sigmoidal shape of the density profiles is indicative of MIPS, at least for large values of $Pe$, and it provides values of the densities of the coexisting stationary states consistent with the maxima observed in the local density probability distribution functions or PDFs (see Appendix).

Subsequent simulations are performed by gradually reducing the value of the P\'eclet number, using as initial configurations the final output of the simulation with the previous value of $Pe$. We continue this procedure until it is no longer possible to clearly identify the coexisting states, which in all the cases considered occurs for $Pe\approx 80$. However, the analysis of the local density PDFs can be used to characterize MIPS up to $Pe\approx 60$.

Under MIPS conditions, in a second stage we considered additional simulations on simulation cells of reduced dimensions $L_x\times L_y=80\times 80$ and density and mole fractions equal to those obtained for the coexisting dense and dilute stationary states obtained in the previous set of the $240\times 40$-cell simulations (i.e. the dense and dilute slabs). In this way, we can characterize the structural and dynamical properties for large homogeneous stationary states which coexist in the MIPS. This is done to eliminate the effect that the presence of the interfaces and their fluctuations may have on the properties of the (homogeneous) coexisting stationary states, which is unavoidable in the $240\times 40$-cell simulations. For the structure of the state, for the last $10^6$ steps of each simulation we monitor the radial distribution functions $g_{\alpha\beta}$ between the $\alpha$ and $\beta$ species and the hexatic order parameter $q_6$. These are defined as
\begin{equation}
g_{\alpha\beta}(r)=\frac{L_x L_y}{N_\alpha N_\beta} \bigg\langle \sum_{i=1}^{N}\sum_{j=1}^{N}\delta_{\alpha\alpha_i}\delta_{\beta\beta_j} (1-\delta_{ij})\;\delta(\mathbf{r}-\mathbf{r}_{ij})\bigg\rangle,   
\end{equation}
where $\alpha_i$ and $\beta_j$ are the species of the particles $i$ and $j$, and $r_{ij}$ is their separation value; and
 \begin{equation}
  \langle q_6 \rangle =\bigg\langle\bigg|\frac{1}{N}  \sum_{j=1}^{N}\sum_{l=1}^{nn_j}\frac{\exp(\iota 6\theta_{jl})}{nn_j}\bigg|\bigg\rangle, 
\end{equation}
where $nn_j$ is of the number of nearest-neighboring particles which are at a distance of the particle $j$ smaller than the cutoff distance $r_c=2^{1/6}$, $\iota$ is the imaginary unit and $\theta_{jl}$ is the angle that the bond between the pair of particles $j$ and $l$ forms with respect to the $x-$axis, which we take as reference axis. This parameter takes values between 0, corresponding to a spatially disordered state, and 1, which characterizes a perfect triangular lattice. 

Finally, the dynamics of the stationary states is characterized through the mean-square displacement (MSD) associated to each species:
\begin{equation}
\langle \Delta r \rangle_\alpha (t)=\frac{1}{N_\alpha}\Bigg\langle\sum_{i=1}^N \delta_{\alpha_i\alpha}|\mathbf{r}_i(t)-\mathbf{r}_i(0)|^2\Bigg\rangle.
\end{equation}
If the system shows diffusive behavior, we find the asymptotic large-$t$ behavior $\langle \Delta r \rangle_\alpha (t)\sim 4D_\alpha t$, where $D_\alpha$ is the self-diffusion coefficient.

\section{\label{results}Results}

\subsection{Phase diagram}
\begin{figure}[t] % [h] = here, [t] = top, [b] = bottom, [p] = página flotante
  \centering
  \includegraphics[width=0.45\textwidth]{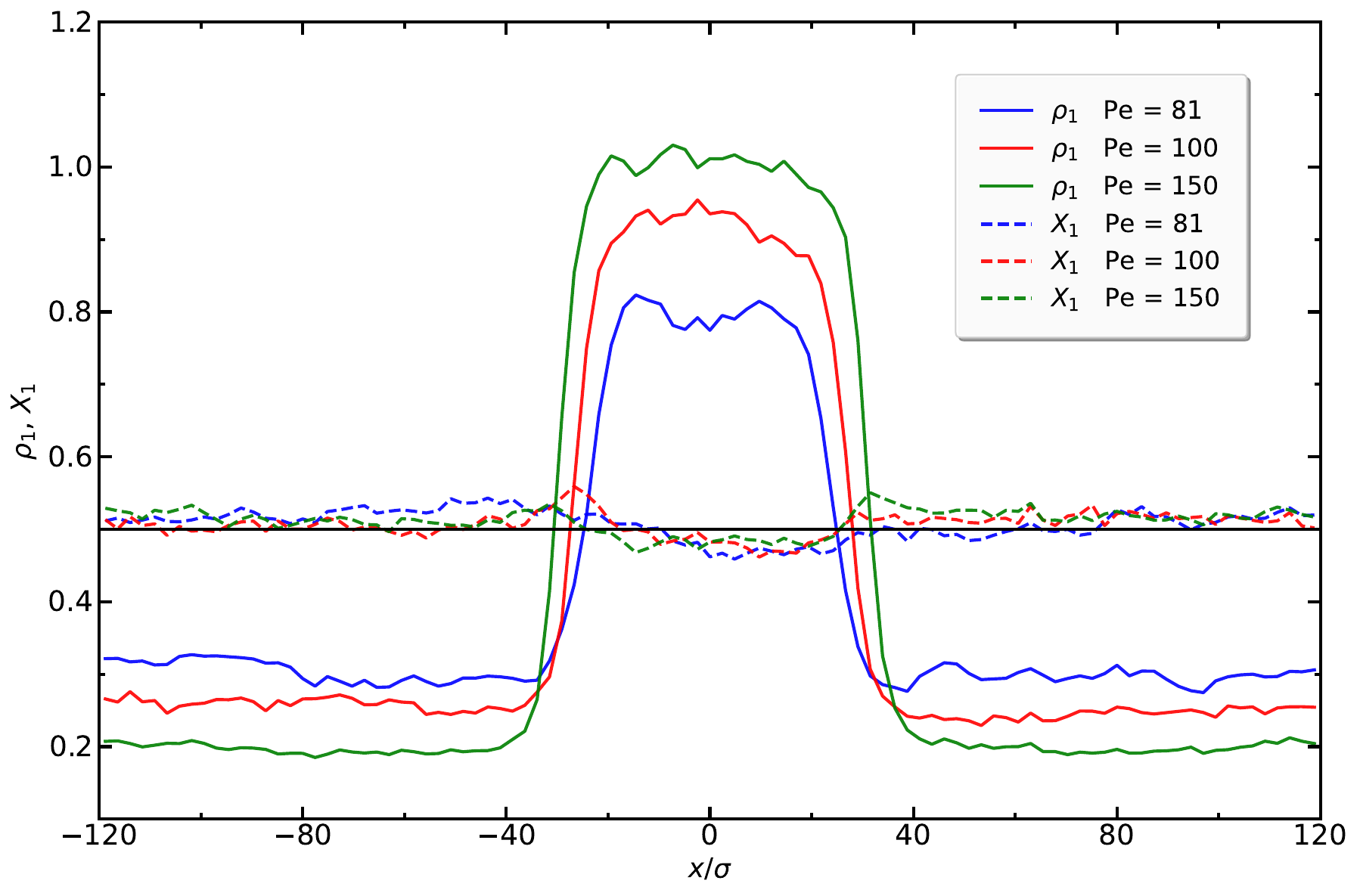} % ruta al archivo
  \hfill
  \includegraphics[width=0.45\textwidth]{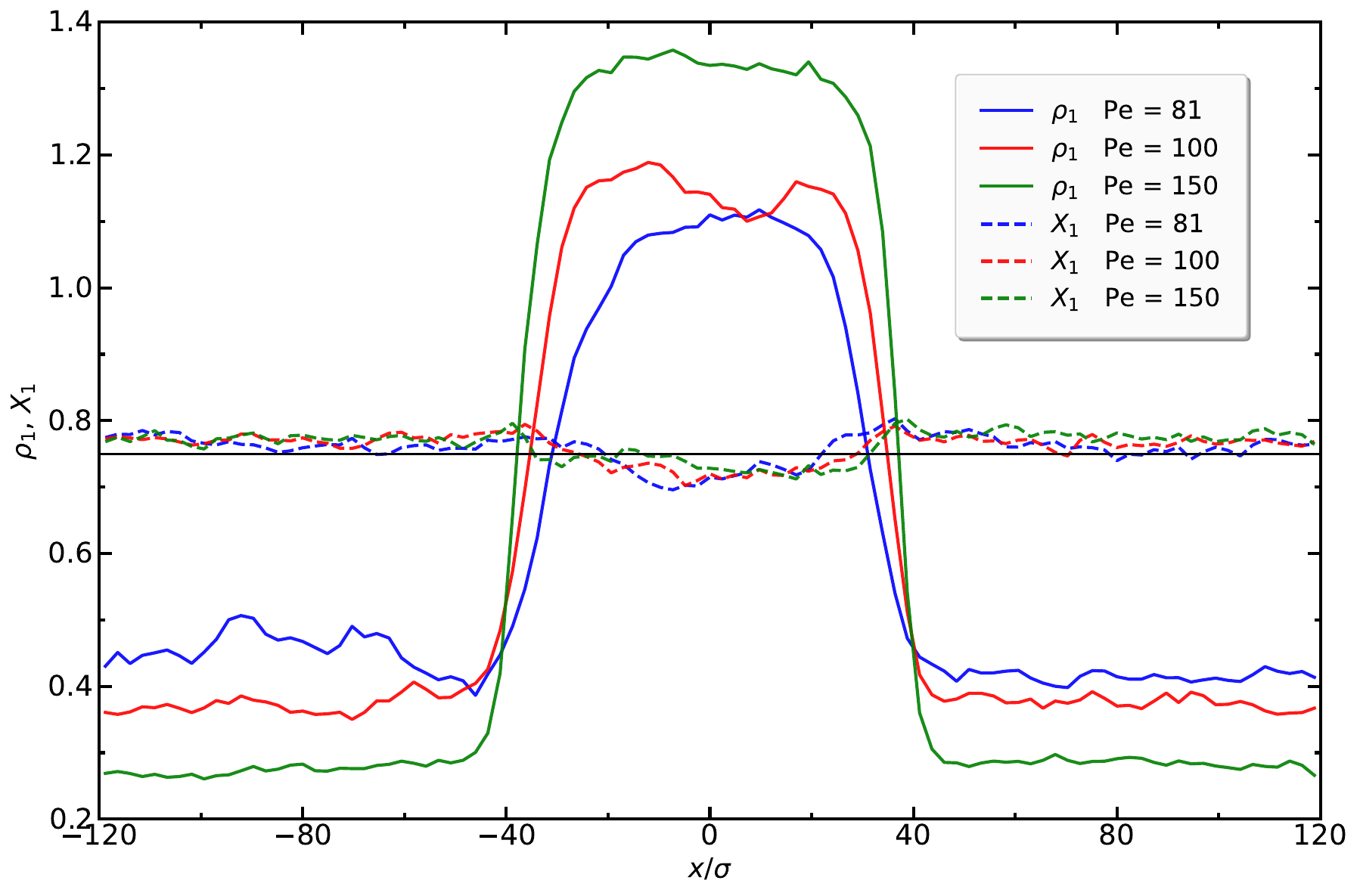} % ruta al archivo
  \caption{Plot of the $1-$component density $\rho_1$ (continuous lines) and mole fraction $X_1$(dashed lines) as a function of the $x-$coordinate for values of the P\'eclet number $Pe=81, 100$ and $150$. For comparison, the value of the global mole fraction is represented by a straight horizontal line. Top figure corresponds to a global mole fraction of $X_1=0.5$ and the bottom figure to the case $X_1=0.75$.}
  \label{fig:molar_fractions}
\end{figure}
As mentioned in the previous Section, MIPS is identified by monitoring the density and mole fraction profiles in our simulations for the elongated box. Fig.~\ref{fig:molar_fractions} represents the typical $1-$component density profile $\rho_1(x)$ and its associated mole component profile $X_1(x)$ for the binary mixture cases given by Eqs. (\ref{rho1profile}) and (\ref{X1profile}), respectively. Note that the $2-$component density $\rho_2(x)$ and total density profile $\rho(x)=\rho_1(x)+\rho_2(x)$ can be obtained from $\rho_1(x)$ and $X_1(x)$ as $\rho_2(x)=\rho_1(x)(1-X_1(x))/X_1(x)$ and $\rho(x)=\rho_1(x)/X_1(x)$. First, we note that the density profile shows a sigmoidal shape, typical of coexistence between stationary states with separating interfaces perpendicular to the $x$ axis. On the other hand, the mole fraction profile is almost homogeneous along the $x-$axis and virtually indistinguishable from the global value. Strictly speaking, a slight decrease in the mole fraction seems to occur in the dense state, although its difference with the corresponding value in the gas-like state is within statistical fluctuations. This observation indicates that particles of the minority class 2 are not segregated and that both $\rho_2(x)$ and $\rho(x)$ are essentially proportional to $\rho_1(x)$. The plateau values of the density profiles determine the densities of a dense and a dilute stationary states in coexistence. For the dense states the density decreases by reducing the P\'eclet number, while the opposite occurs for the dilute states. For $Pe<80$, density fluctuations increase and our density profiles analysis does not provide a reliable method to obtain the densities of the stationary coexisting states. However, the analysis of the density PDFs and, in particular, its bimodal shape, allows us to extend the coexistence curve to values of $Pe\approx 60$, indicating that the critical point for the MIPS is about $Pe=60-70$ for all the mixtures considered in this work. However, in our work we do not pursue a precise location of the critical point. 

\begin{figure}[t] % [h] = here, [t] = top, [b] = bottom, [p] = página flotante
  \centering
  \includegraphics[width=0.45\textwidth]{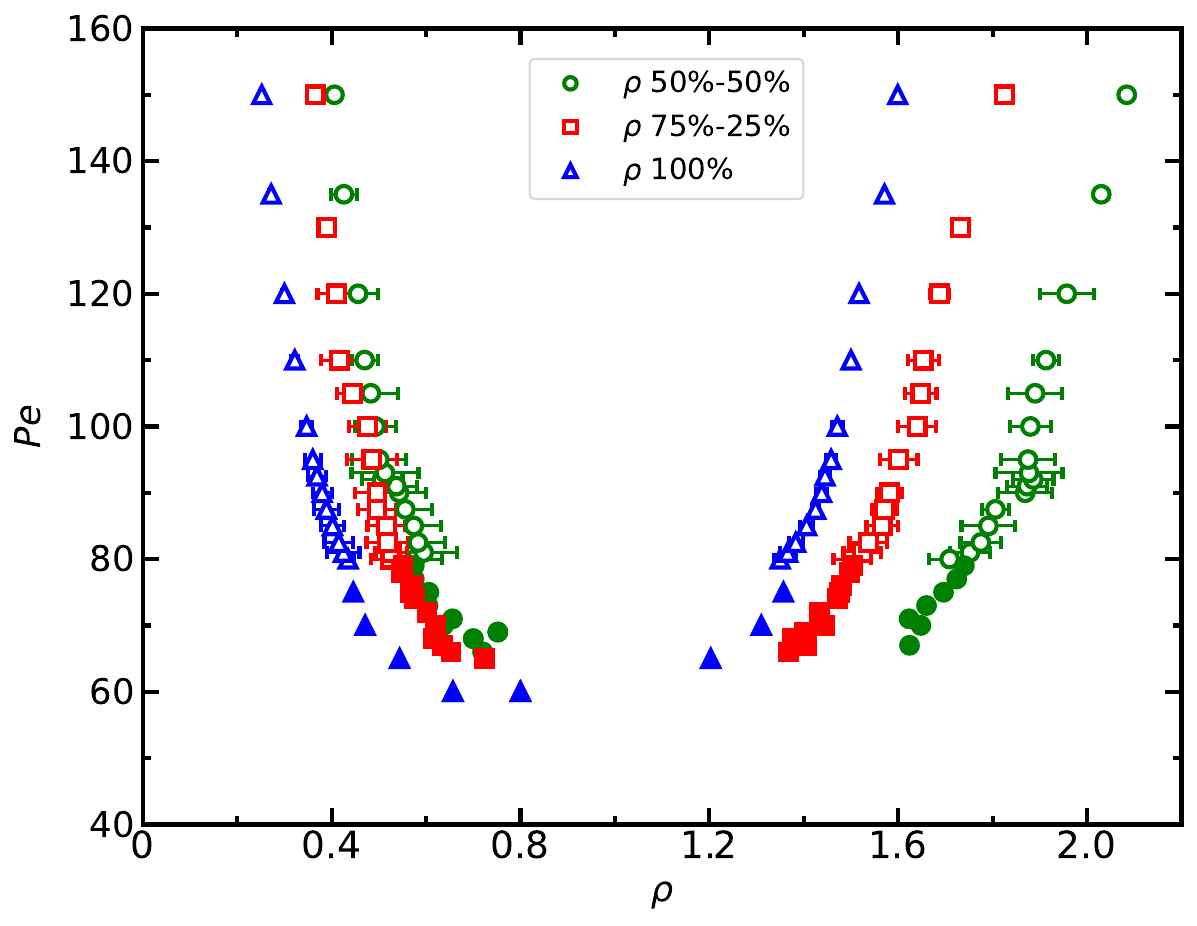} % ruta al archivo
  \caption{MIPS $\rho-Pe$ phase diagram for global mole fractions $X_1=1$ (blue triangles), $0.75$ (red squares) and $0.5$ (green circles). Open symbols are the densities of the coexisting stationary states obtained from the analysis of the density profiles, and the filled symbols correspond to the values obtained from the density PDFs.}
  \label{fig:phase_diagram}
\end{figure}

Fig.\ref{fig:phase_diagram} shows the densities of the coexisting stationary states as a function of the P\'eclet number for the monocomponent system and the binary mixtures with mole fractions $X_1=0.75$ and $0.5$. Although the the critical P\'eclet number seems to be insensitive to the composition of the mixture, the system mole fraction has a strong effect on the densities of the coexisting states. We note that the densities of the coexisting dilute and dense phases in the monocomponent system are systematically smaller than those of the binary systems. On the other hand, the binary mixtures show very similar densities along the dilute states branch, but the higher the mole fraction $X_1$, the larger the density for the dense states branch. This fact indicates that there is no corresponding state relationship as the coexisting phases do not collapse when rescaled by the critical value. 

\begin{figure}[t] % [h] = here, [t] = top, [b] = bottom, [p] = página flotante
  \centering
  \includegraphics[width=0.45\textwidth]{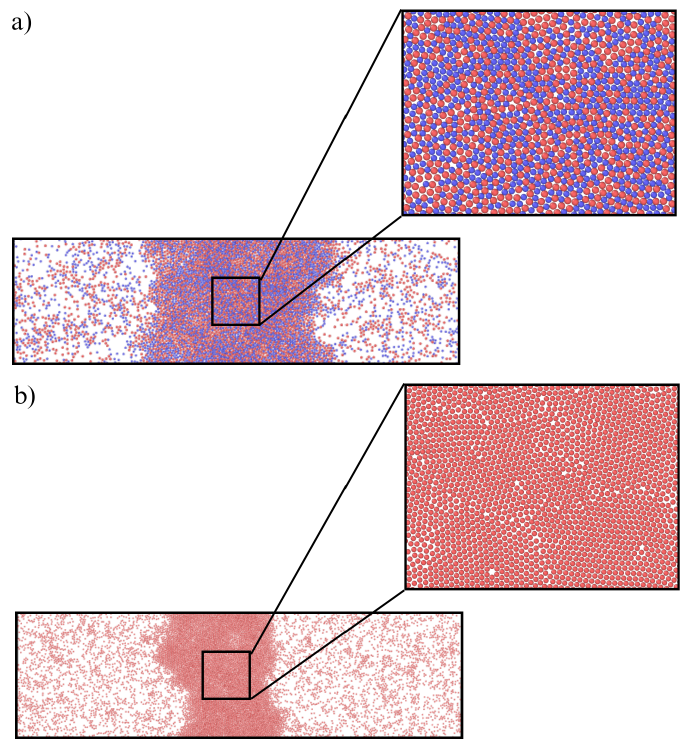} % ruta al archivo
  \caption{Snapshot of the $240\times 40$-cell simulation for $Pe=100$ for $X_1=0.5$ (a) and $X_1=1$ (b). Red disks correspond to $1-$type particles and blue disks to $2-$type particles. A region on the high-density regions is zoomed.}
  \label{fig:snapshot}
\end{figure}

Although MIPS phase diagrams seem to be qualitatively similar for all the considered cases, inspection of typical snapshots shows differences between the monocomponent and the binary systems in the high-density phase. Fig.~\ref{fig:snapshot} shows snapshots of the monocomponent system and the binary mixture with $X_1=0.5$ for $Pe=100$. The dense phase in the monocomponent system shows positional ordering, compatible with an hexatic or solid phase, while in the binary mixture it presents liquid-like disorder. Similar results are observed for $X_1=0.75$, confirming that the even a small amount of size polydispersity is sufficient to break down the crystalline order. However, this fact does not significantly affect the occurrence of MIPS, beyond the shift in the densities of the coexisting states.

\subsection{Structural properties of the coexisting stationary states}

\begin{figure*}[t] % [t] o [b]; en dos columnas, figure* sólo va arriba/abajo
  \centering

  % ----- Fila 1 -----
  \begin{subfigure}{0.48\textwidth}
    \centering
    \includegraphics[width=\linewidth]{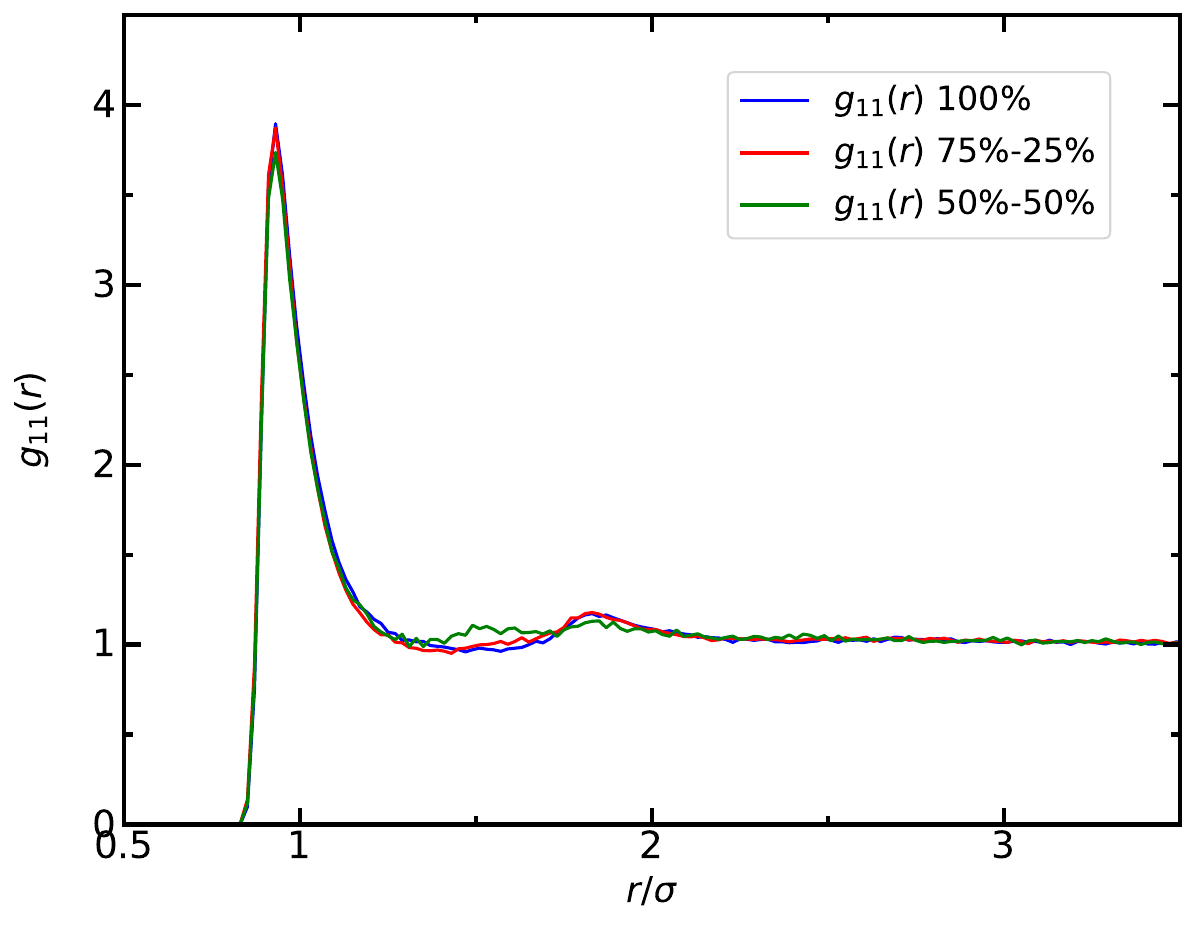}
    %\caption{Título 1}
    %\label{fig:g1}
  \end{subfigure}\hfill
  \begin{subfigure}{0.48\textwidth}
    \centering
    \includegraphics[width=\linewidth]{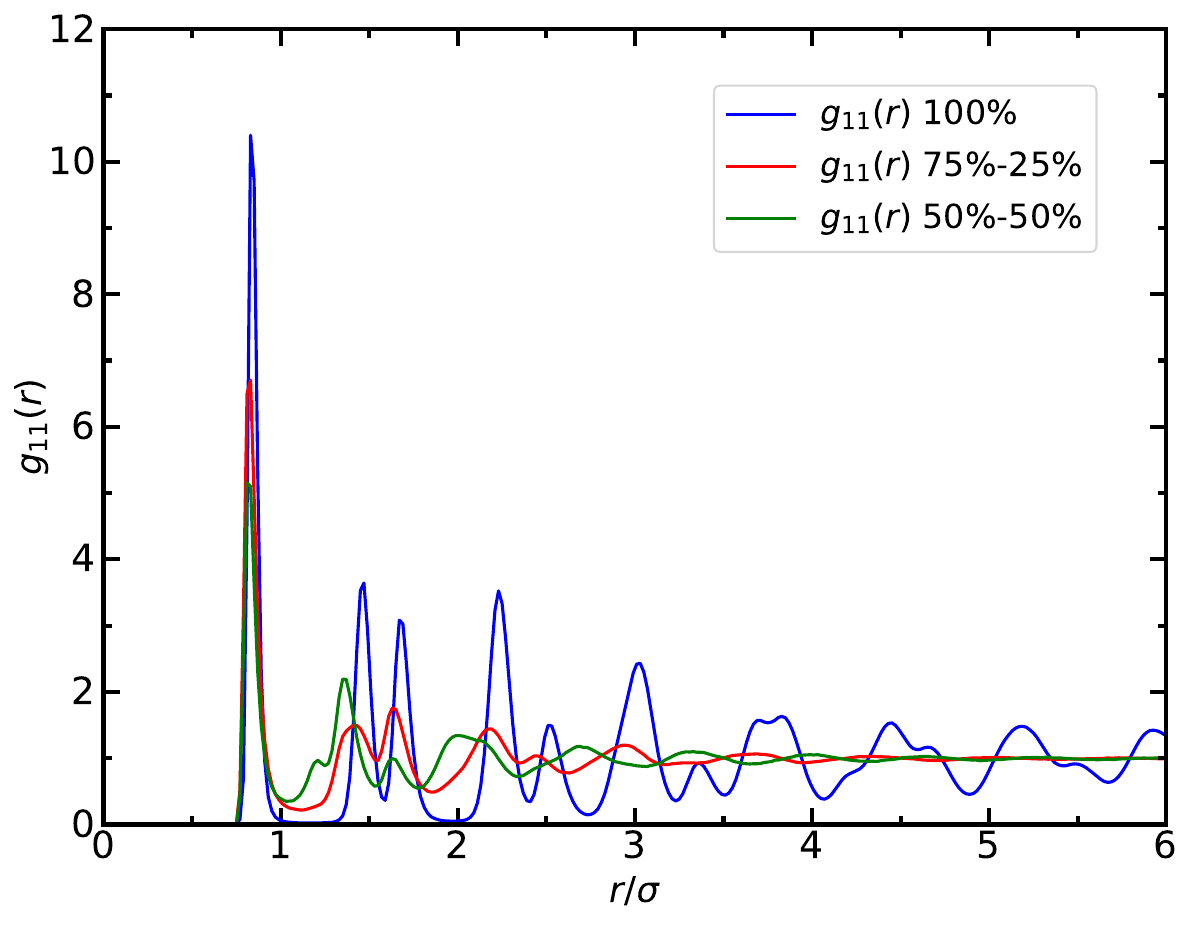}
    %\caption{Título 2}
    %\label{fig:g2}
  \end{subfigure}

  %\vspace{0.2ex}

  % ----- Fila 2 -----
  \begin{subfigure}{0.48\textwidth}
    \centering
    \includegraphics[width=\linewidth]{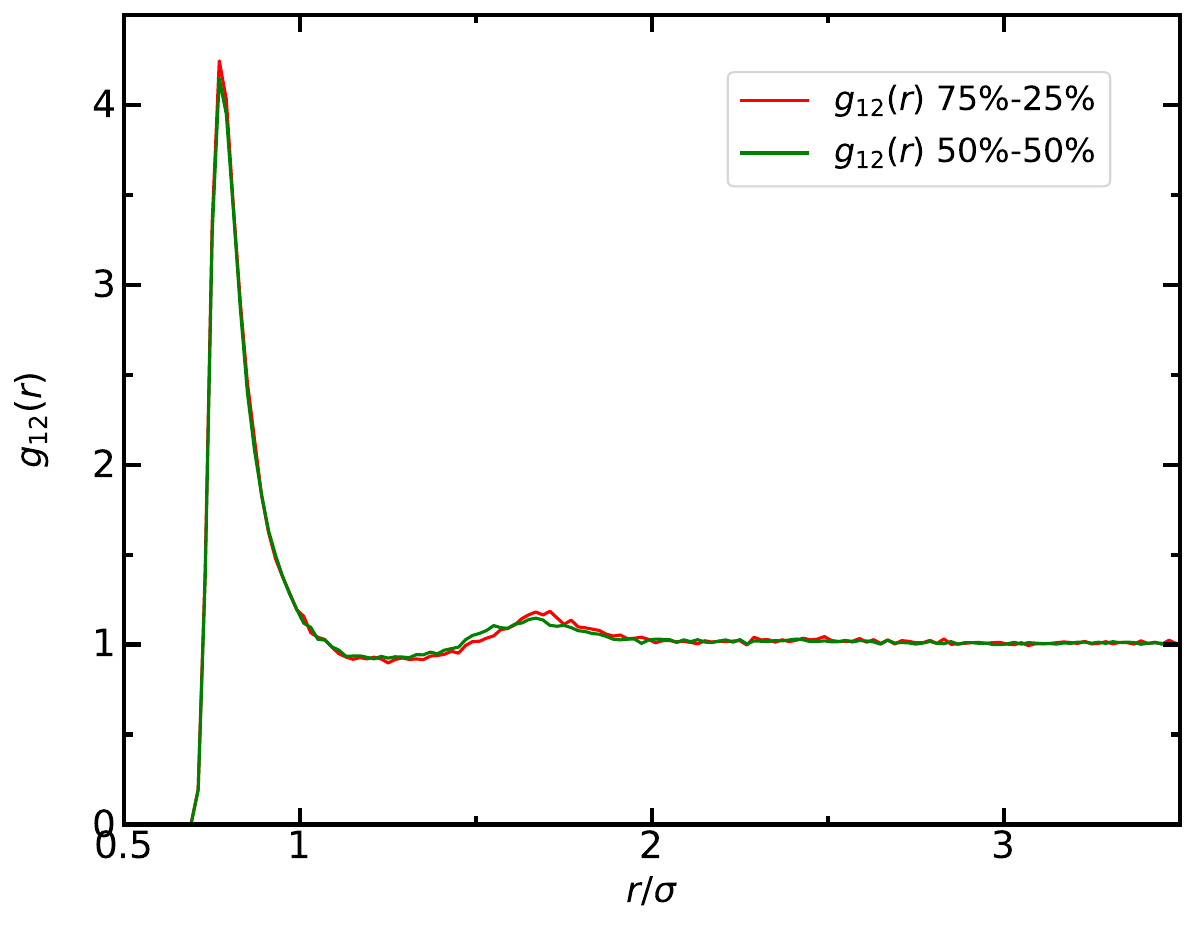}
   % \caption{Título 3}
    %\label{fig:g3}
  \end{subfigure}\hfill
  \begin{subfigure}{0.48\textwidth}
    \centering
    \includegraphics[width=\linewidth]{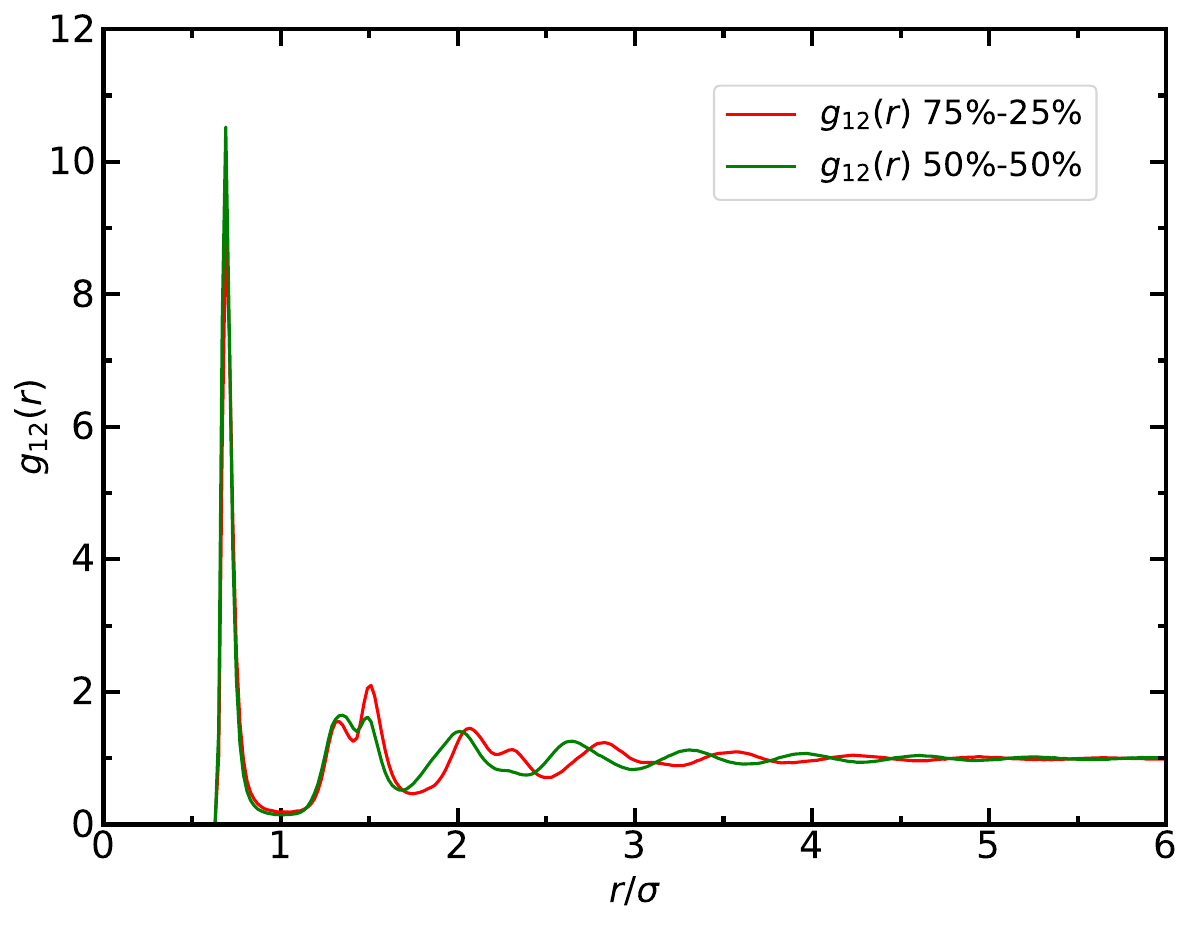}
  %  \caption{Título 4}
    %\label{fig:g4}
  \end{subfigure}

  %\vspace{0.2ex}

  % ----- Fila 3 -----
  \begin{subfigure}{0.48\textwidth}
    \centering
    \includegraphics[width=\linewidth]{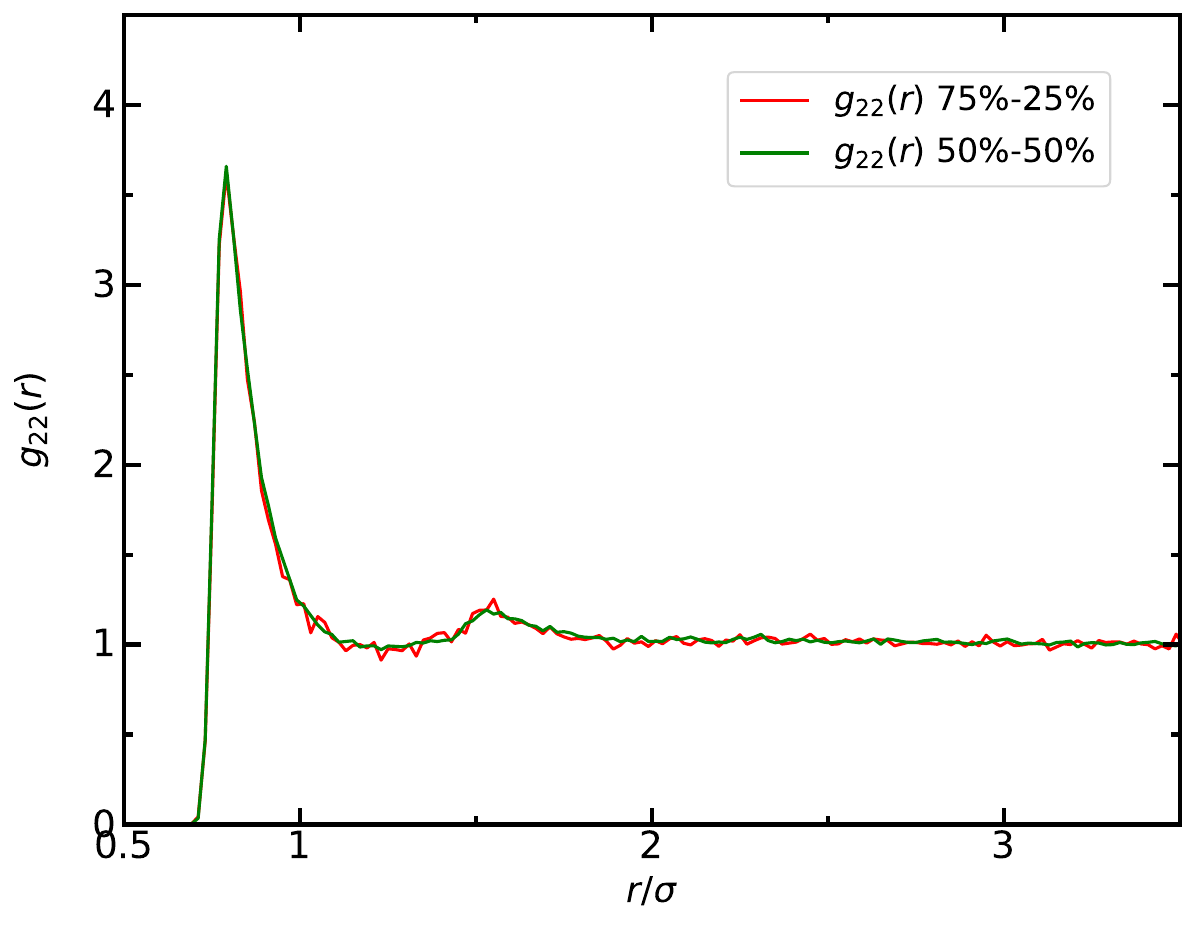}
 %   \caption{Título 5}
    %\label{fig:g5}
  \end{subfigure}\hfill
  \begin{subfigure}{0.48\textwidth}
    \centering
    \includegraphics[width=\linewidth]{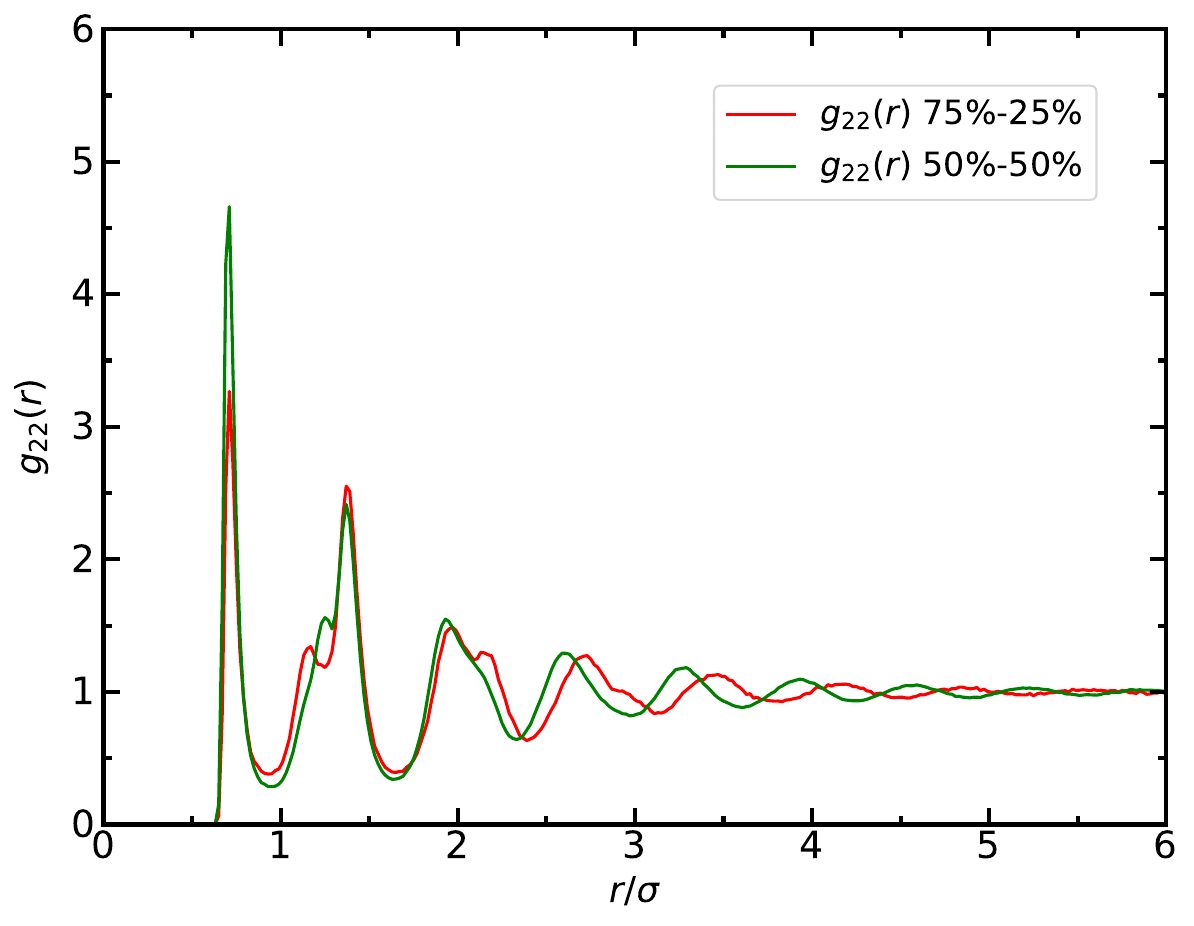}
   % \caption{Título 6}
    %\label{fig:g6}
  \end{subfigure}

   % \vspace{0.2ex}

  \caption{Radial distribution functions $g_{11}, g_{12}$ and $g_{22}$ of the coexisting stationary states for $Pe=150$. The left column corresponds to the dilute state, and the right to the high-density state.}
  \label{fig:gr}
\end{figure*}

In order to characterize the structure of the coexisting stationary states, we first obtained the radial distribution functions for each state. We observe that these quantities are essentially independent of the P\'eclet number. Thus, we will restrict ourselves to radial distribution functions for the $Pe=150$ case, which are plotted in Fig.~\ref{fig:gr}. For other P\'eclet numbers, the features of the radial distribution functions are qualitatively similar. For the dilute phases, all the radial distribution functions are almost independent of the mole fraction, showing a peak associated to the effective collision diameter between the particles of the pair and a shallow secondary minimum at about twice the first peak position, decaying to 1 for large separation distances. The effective collision diameter between a pair of particles of species $\alpha$ and $\beta$ is usually smaller than the value $\sigma_{\alpha\beta}$, indicating the softness of the potential. In general, they are slightly decreasing with the P\'eclet number: for the monocomponent system, it is approximately equal to $0.96\pm 0.1$, whereas for both binary mixtures the effective collision diameter between like $1-$type particles, unlike particles and like $2-$type particles span over ranges $0.93-0.96$, $0.77-0.80$ and $0.79-0.82$, respectively. 

Now we turn to the radial distribution functions for the high-density phase. The first peak is again almost independent of the mole fraction. However, at larger separations more differences are observed, as expected. In the radial distribution function for $1-$type like particles, the monocomponent system shows a behavior compatible with a solid-like ordering, while the binary cases show a dense-liquid-like behavior with damped oscillations which decay to $1$ for large separation distances. However, the second and third coordination shells present some structure which is reminiscent of a short-ranged local ordering. This is more pronounced for the case $X_1=0.75$ than for $X_1=0.5$. On the other hand, the radial distribution functions between unlike particles and $2-$type like particles for the binary cases are more similar between them and show less structure in the second and third coordination shells than in $g_{11}$.  

\begin{figure}[t] % [h] = here, [t] = top, [b] = bottom, [p] = página flotante
  \centering
  \includegraphics[width=0.45\textwidth]{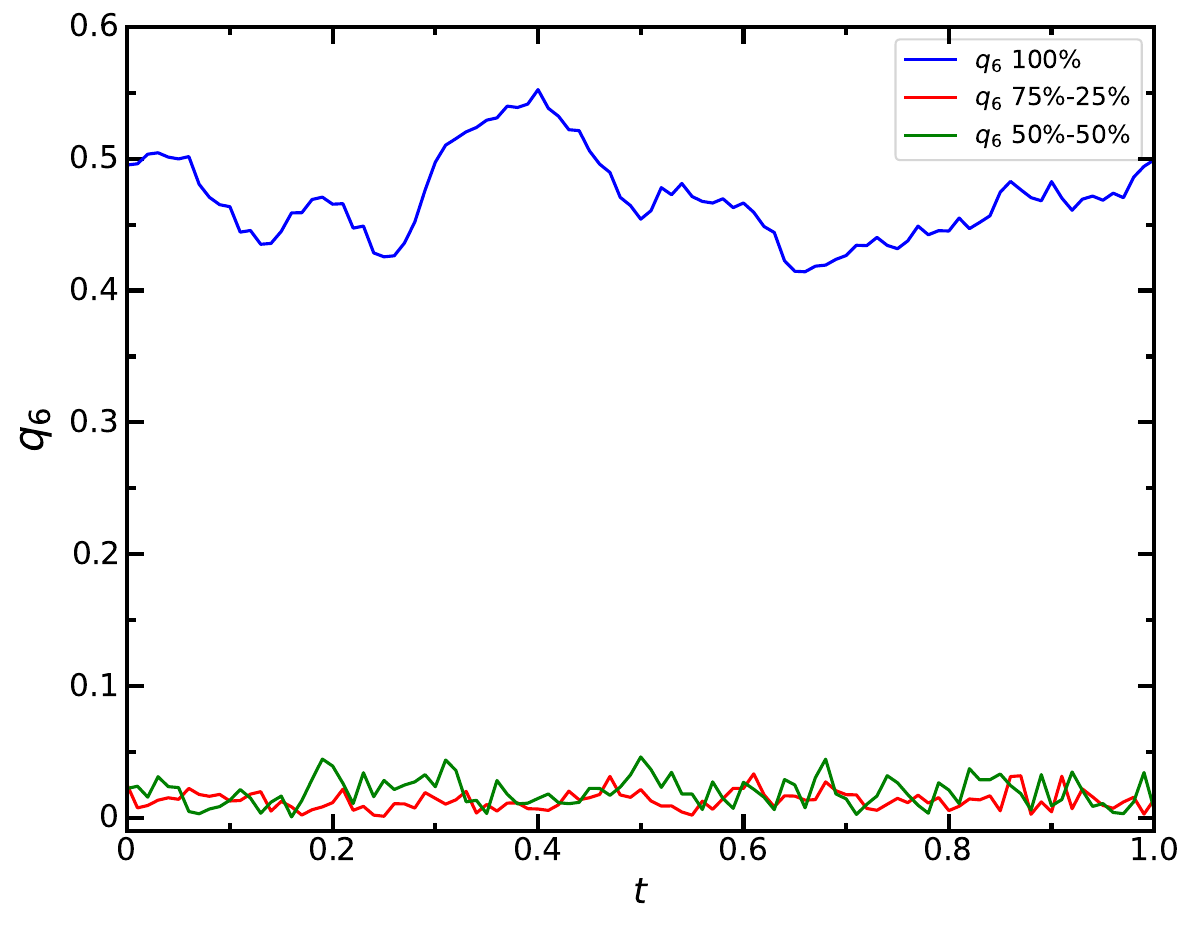} % ruta al archivo
  \hfill
  \includegraphics[width=0.45\textwidth]{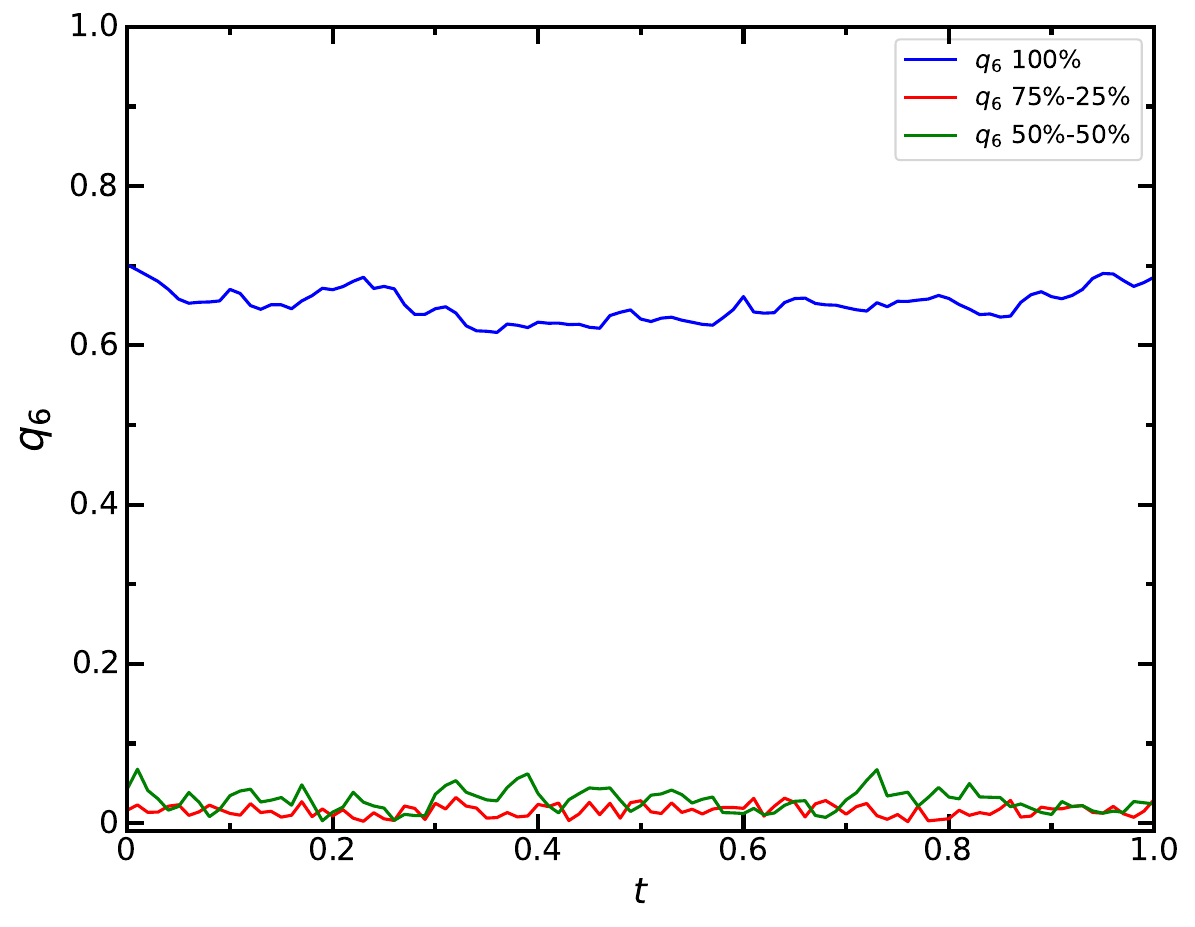} % ruta al archivo
  \hfill
  \includegraphics[width=0.45\textwidth]{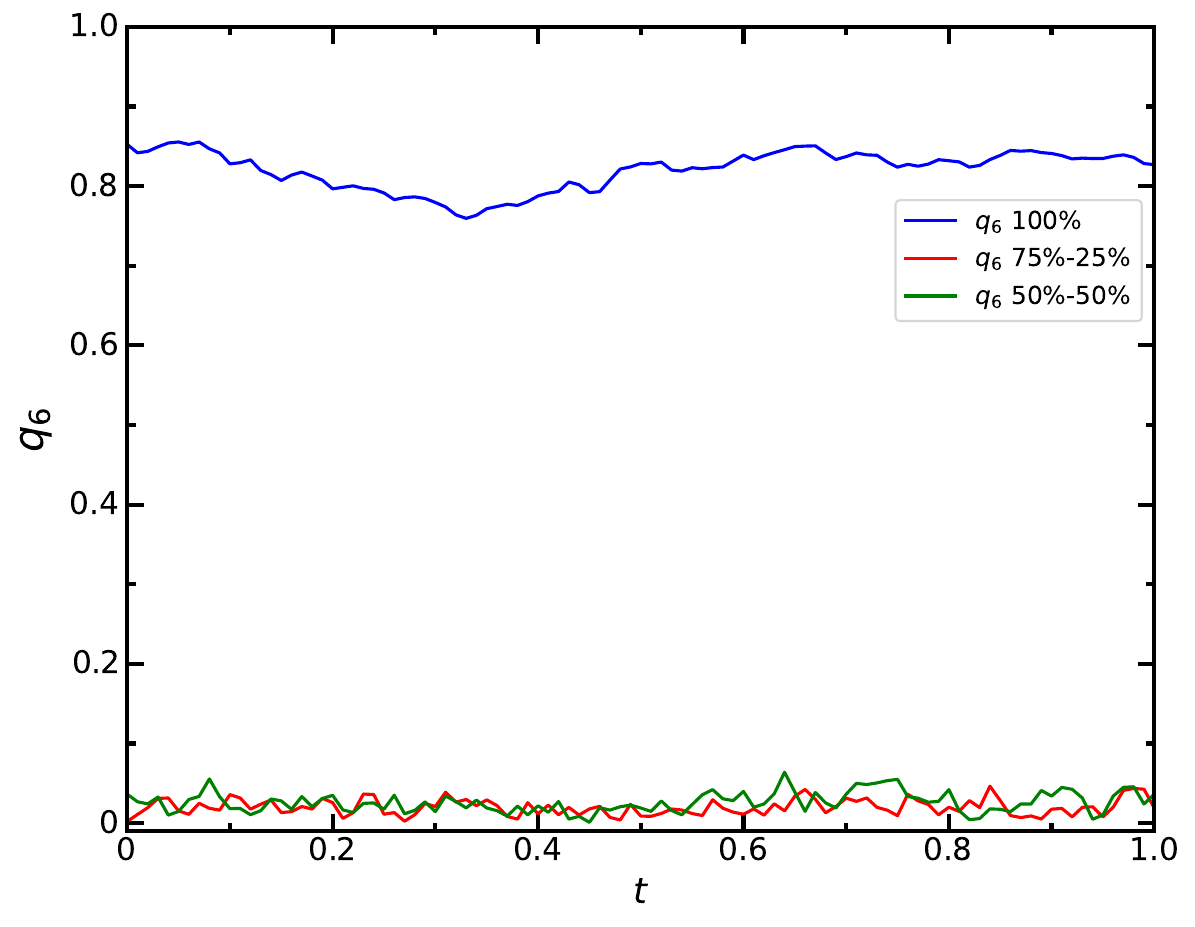} % ruta al archivo
  
  \caption{Instantaneous hexatic order parameter over the last $10^6$ time steps for $X_A=1$ (blue line), $0.75$ (red line) and $0.5$ (green line). From top to bottom: $Pe = 81, 100$ and $150$.}
  \label{fig:hexatic_order}
\end{figure}

Now we turn to the analysis of the hexatic order parameter $\langle q_6\rangle$. We restrict ourselves to the high-density states. First, we note that the hexatic ordering, when present, is higher in the $80\times 80$-cell simulations of the homogeneous dense phase than in the dense slab for the $240\times 40$-cell simulations. This can be rationalized by the disturbance that introduces the presence of the fluctuating interface in the latter. In agreement with our previous observations, binary systems have negligible values of the hexatic order parameter. On the contrary, the monocomponent system shows a non-vanishing value of $\langle q_6\rangle$ over all the range of P\'eclet numbers in which MIPS appears, in agreement with previous simulation studies \cite{CARTV22}. The value of $\langle q_6\rangle$ decreases as the P\'eclet number is reduced and, in addition, it fluctuates in a wider range of values, as it is shown in Fig.\ref{fig:hexatic_order}. In particular, the plot for $Pe=81$ in the monocomponent case suggests a near-critical character of this state. 

\subsection{Mean-square displacement}

\begin{table*}
\caption{\label{tab:DIFF} Self-diffusion coefficients estimates in the coexisting states.}
\begin{ruledtabular}
\begin{tabular}{cccccccccc}
 DILUTE STATE&\multicolumn{3}{c}{$X_1=1$}&\multicolumn{3}{c}{$X_1=0.75$}&\multicolumn{3}{c}{$X_1=0.5$}\\ \hline
 
 P\'eclet number &$Pe=81$&$Pe=100$&$Pe=150$&$Pe=81$&$Pe=100$&$Pe=150$&$Pe=81$&$Pe=100$&$Pe=150$ \\
 
$D_1$& $5.7\times 10^2$ & $8.9\times 10^2$ & $3.2\times 10^3$ & $4.0\times 10^2$&$6.7\times 10^2$&$1.5\times 10^3$&$5.3\times 10^2$&$9.2\times 10^2$&$3.1\times 10^3$ \\
$D_2$&-&-&-&$4.7\times 10^2$&$8.6\times 10^2$&$1.6\times 10^3$&$5.8\times 10^2$&$9.0\times 10^2$&$2.4\times 10^3$\\
Ideal ABP &$1.1\times 10^3$&$1.7\times 10^3 $&$3.7\times 10^3$&$1.1\times 10^3$&$1.7\times 10^3 $&$3.7\times 10^3$&$1.1\times 10^3$&$1.7\times 10^3 $&$3.7\times 10^3$ \\
\hline\\
DENSE STATE&\multicolumn{3}{c}{$X_1=1$}&\multicolumn{3}{c}{$X_1=0.75$}&\multicolumn{3}{c}{$X_1=0.5$}\\ \hline
 
 P\'eclet number &$Pe=81$&$Pe=100$&$Pe=150$&$Pe=81$&$Pe=100$&$Pe=150$&$Pe=81$&$Pe=100$&$Pe=150$ \\
 
$D_1$& $9.3$ & $1.5$ & $3.1$ & $18$&$18$&$24$&$21$&$18$&$19$ \\
$D_2$&-&-&-&$21$&$19$&$21$&$19$&$22$&$24$\\

\end{tabular}
\end{ruledtabular}
\end{table*}

\begin{figure*}[t] % [t] o [b]; en dos columnas, figure* sólo va arriba/abajo
  \centering

  % ----- Fila 1 -----
  \begin{subfigure}{0.48\textwidth}
    \centering
    \includegraphics[width=\linewidth]{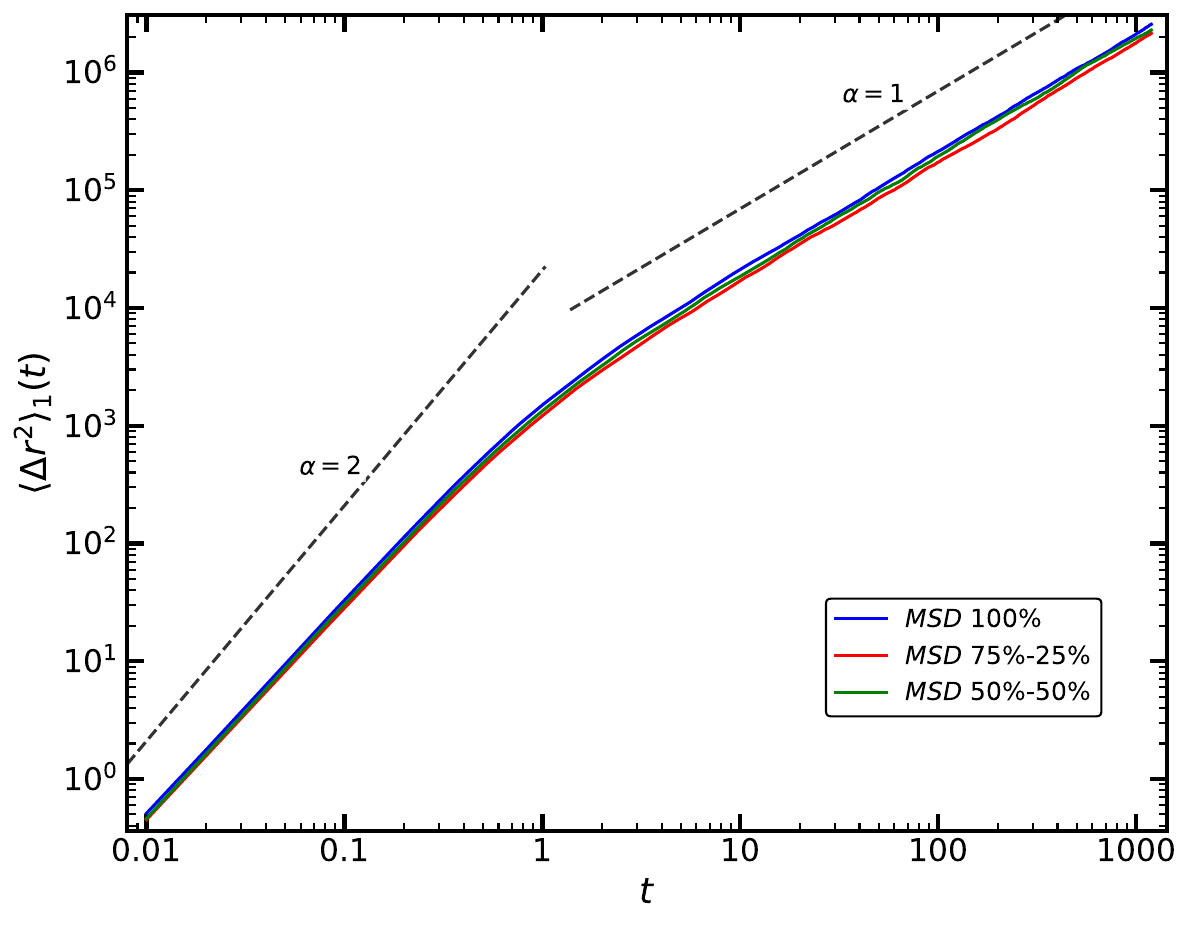}
    %\caption{Título 1}
    \label{fig:msd1}
  \end{subfigure}\hfill
  \begin{subfigure}{0.48\textwidth}
    \centering
    \includegraphics[width=\linewidth]{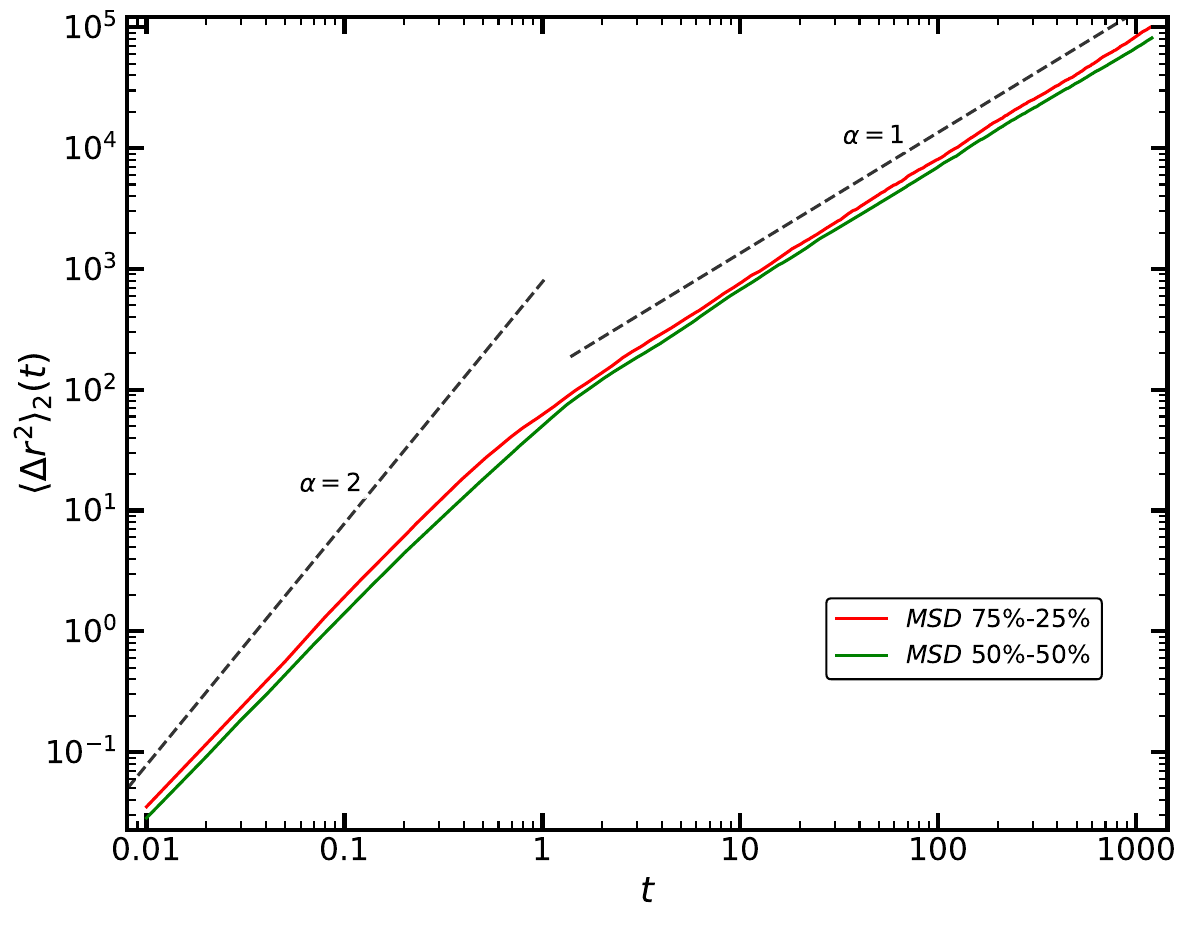}
    %\caption{Título 2}
    \label{fig:msd2}
  \end{subfigure}

  %\vspace{0.2ex}

  % ----- Fila 2 -----
  \begin{subfigure}{0.48\textwidth}
    \centering
    \includegraphics[width=\linewidth]{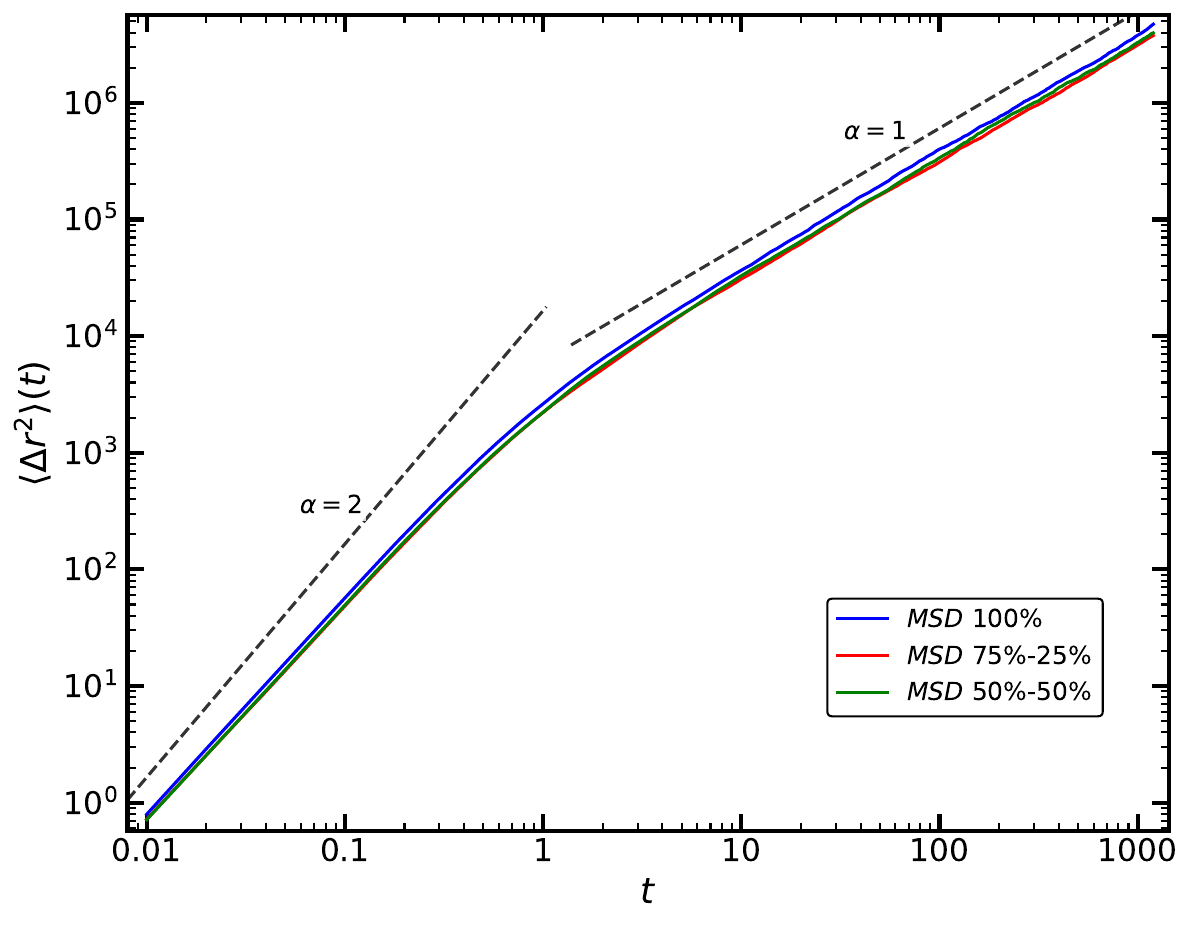}
   % \caption{Título 3}
    \label{fig:msd3}
  \end{subfigure}\hfill
  \begin{subfigure}{0.48\textwidth}
    \centering
    \includegraphics[width=\linewidth]{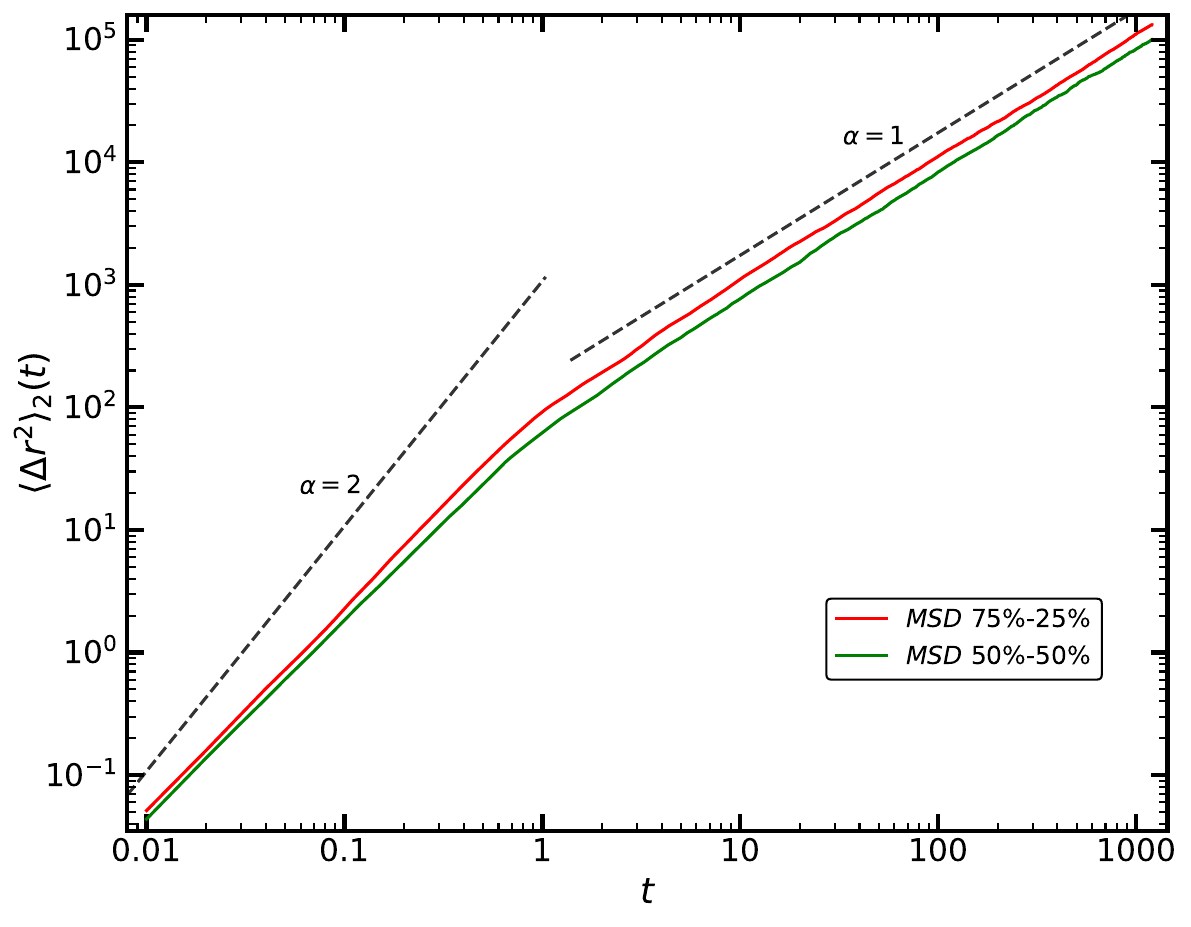}
  %  \caption{Título 4}
    \label{fig:msd4}
  \end{subfigure}

  %\vspace{0.2ex}

  % ----- Fila 3 -----
  \begin{subfigure}{0.48\textwidth}
    \centering
    \includegraphics[width=\linewidth]{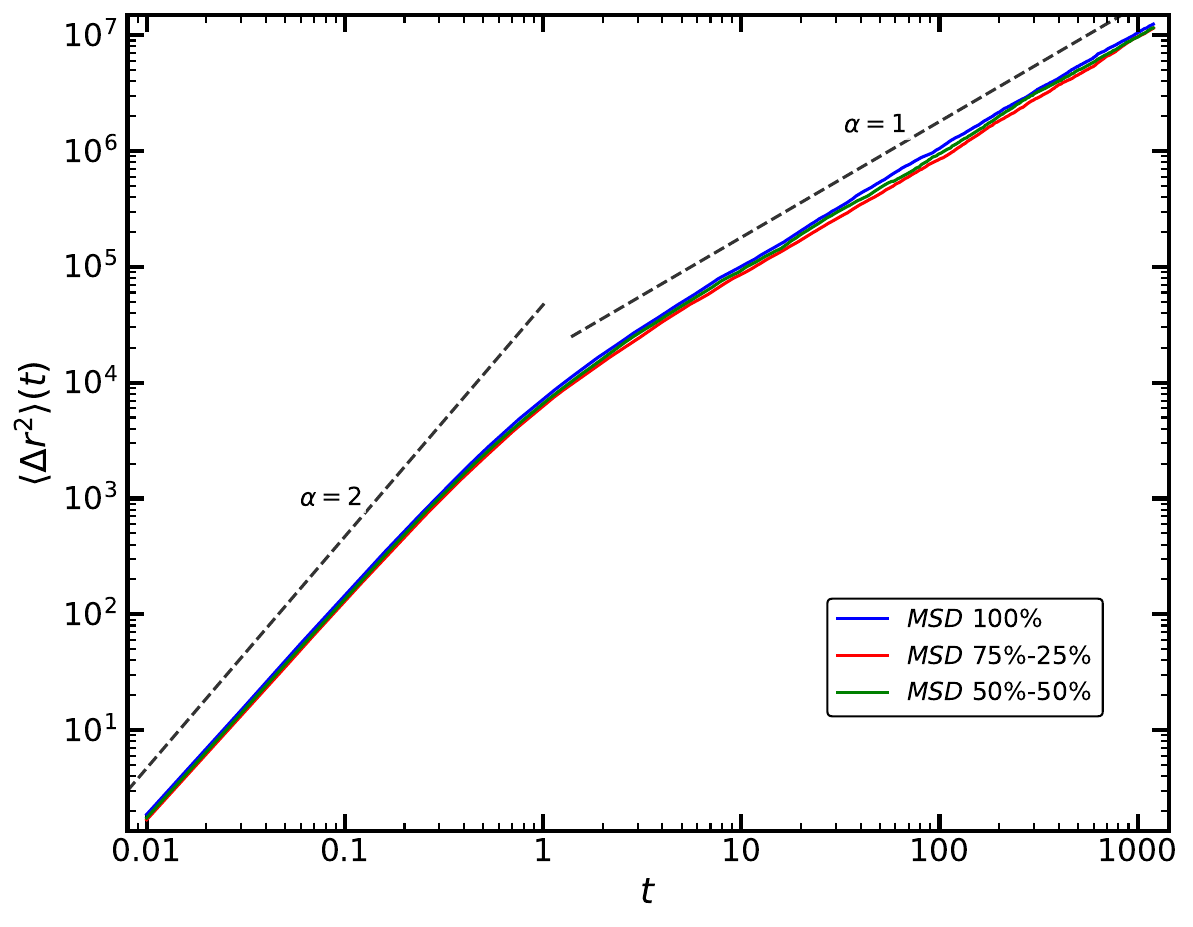}
 %   \caption{Título 5}
    \label{fig:msd5}
  \end{subfigure}\hfill
  \begin{subfigure}{0.48\textwidth}
    \centering
    \includegraphics[width=\linewidth]{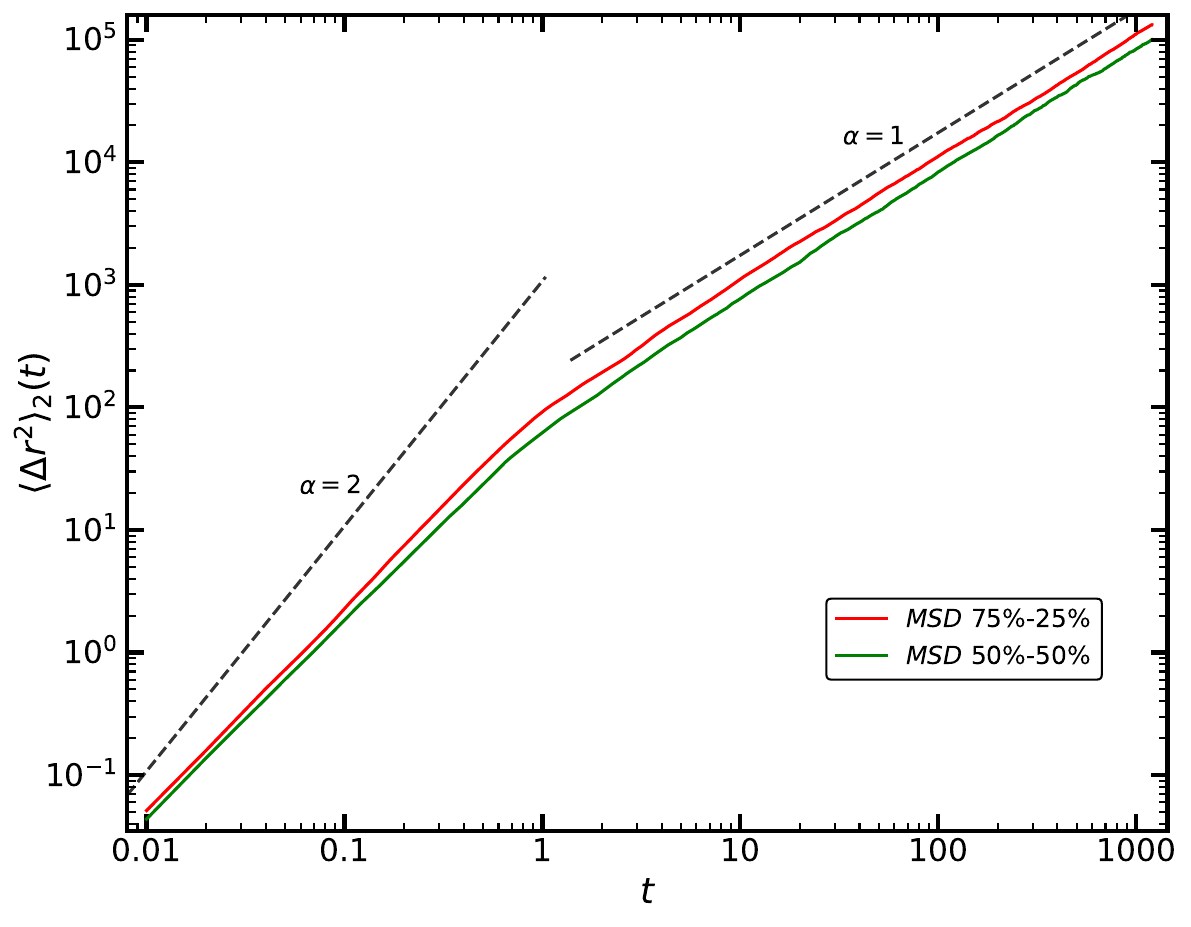}
   % \caption{Título 6}
    \label{fig:msd6}
  \end{subfigure}

  \caption{Log-log plot of the mean-square displacements as a function of time in the dilute coexisting state. In each plot, we represented the results for $X_1=1$ (blue lines), $0.75$ (red lines) and $0.25$ (green lines). The left column corresponds to the MSD of $1-$type particles and the right column to the MSD of $2-$type particles. From top to bottom: $Pe = 81, 100$ and $150$.}
  \label{fig:msd}
\end{figure*}

The results described in the previous subsection are independent of time. For example, they cannot distinguish between liquid-like or glassy amorphous states in disordered cases. Thus, we have to analyze dynamical properties to discern between these options. In this subsection we analyze the mean-square displacement of the particles in both coexisting states for $Pe=81, 100$ and $150$. For the dilute state, the MSD shows a smooth crossover from a short-time ballistic behavior (in which $\langle (\Delta r)^2\rangle\sim t^2$) to a diffusive behavior $\langle (\Delta r)^2\rangle\sim 4D t$ for large $t$ (see Fig.\ref{fig:msd}). Self-diffusion coefficients can be estimated from the latter (see Table \ref{tab:DIFF}). Although the MSD plots seem to be quite insensitive to the composition for both species and P\'eclet number, they can differ by a factor of at most 2. On the other hand, diffusion coefficients for the $2-$type particles are systematically larger than for species $1$. It is instructive to compare these estimates with the ideal ABP prediction \cite{Winkler}, in which the self-diffusion coefficient is given by
\begin{equation}
D=D_S+\frac{v_a^2}{2D_R}=D_S\left(1+\frac{Pe^2}{6}\right),
\label{idealabp}
\end{equation}
regardless of the species since their dynamics is identical in the absence of interactions between particles. We note that, in the range of P\'eclet numbers in which MIPS appears, the ideal self-diffusion coefficient is dominated by activity, so the higher the value of $Pe$, the higher the diffusion coefficient. Although there is no quantitative agreement between our estimations of the diffusion coefficientes and the ideal ABP predictions (which indicates that interactions have an effect on diffusion coefficients even in the dilute coexisting state), they capture the correct order of magnitude (up to a factor of at most 2) and its qualitative behavior with the P\'eclet number.  

\begin{figure*}[t] % [t] o [b]; en dos columnas, figure* sólo va arriba/abajo
  \centering

  % ----- Fila 1 -----
  \begin{subfigure}{0.48\textwidth}
    \centering
    \includegraphics[width=\linewidth]{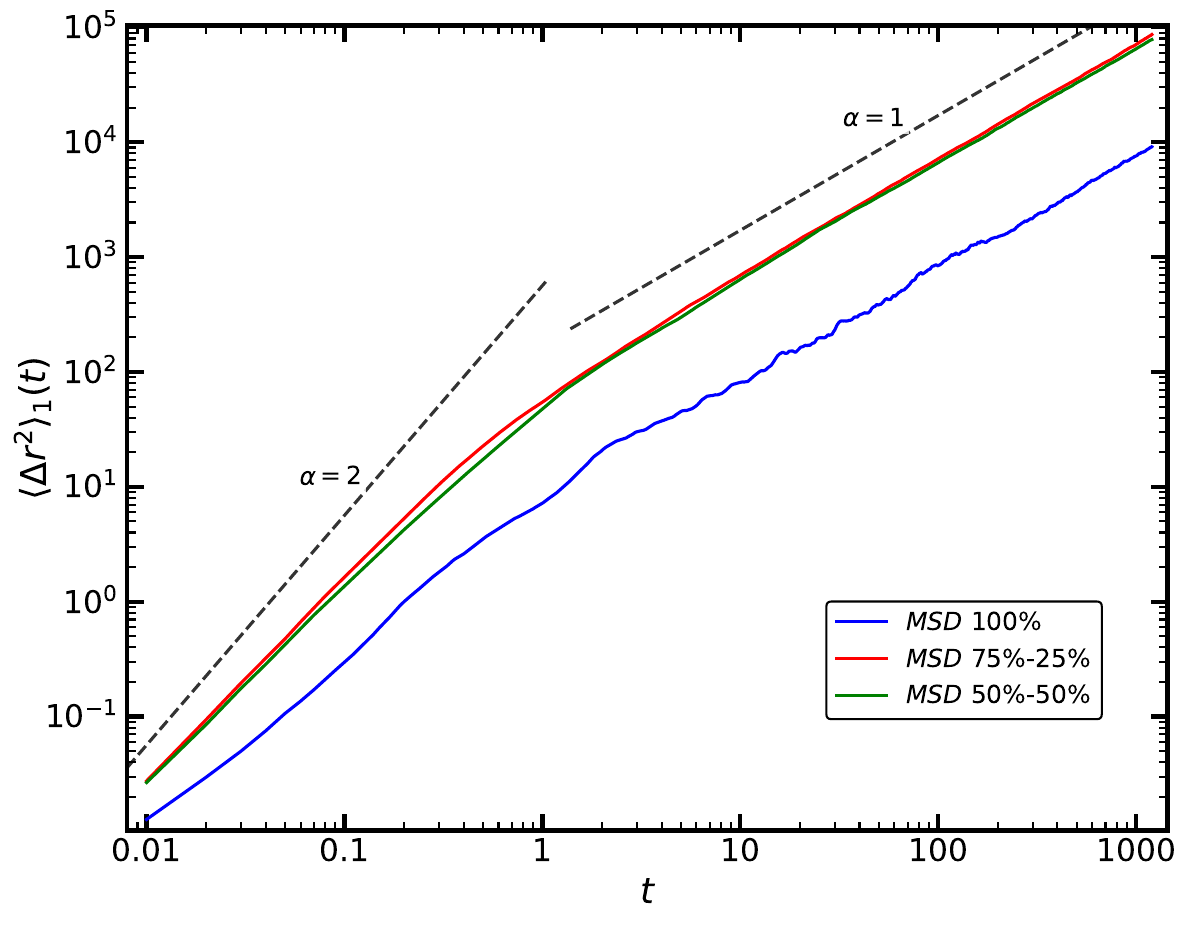}
    %\caption{Título 1}
    %\label{fig:msd1}
  \end{subfigure}\hfill
  \begin{subfigure}{0.48\textwidth}
    \centering
    \includegraphics[width=\linewidth]{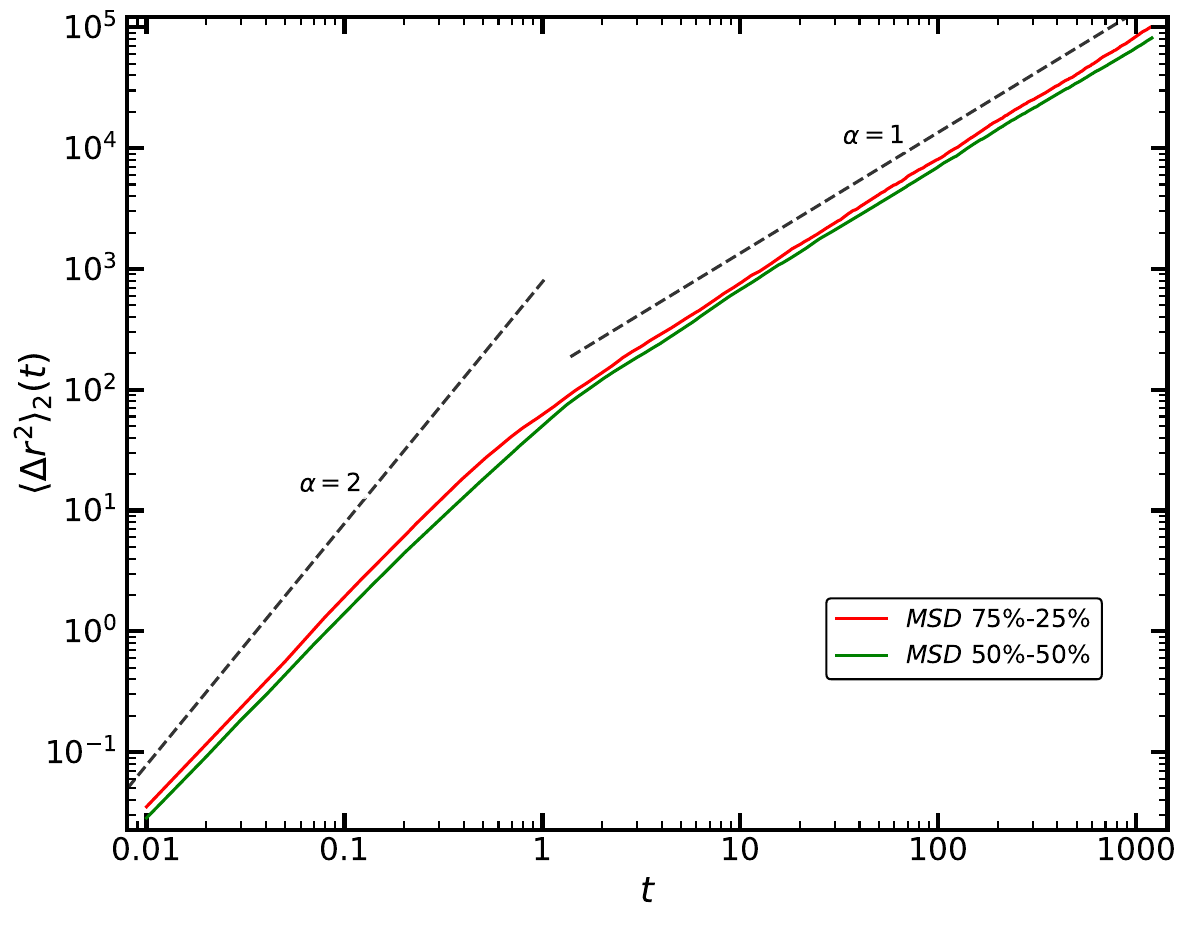}
    %\caption{Título 2}
    %\label{fig:msd2}
  \end{subfigure}

  %\vspace{0.2ex}

  % ----- Fila 2 -----
  \begin{subfigure}{0.48\textwidth}
    \centering
    \includegraphics[width=\linewidth]{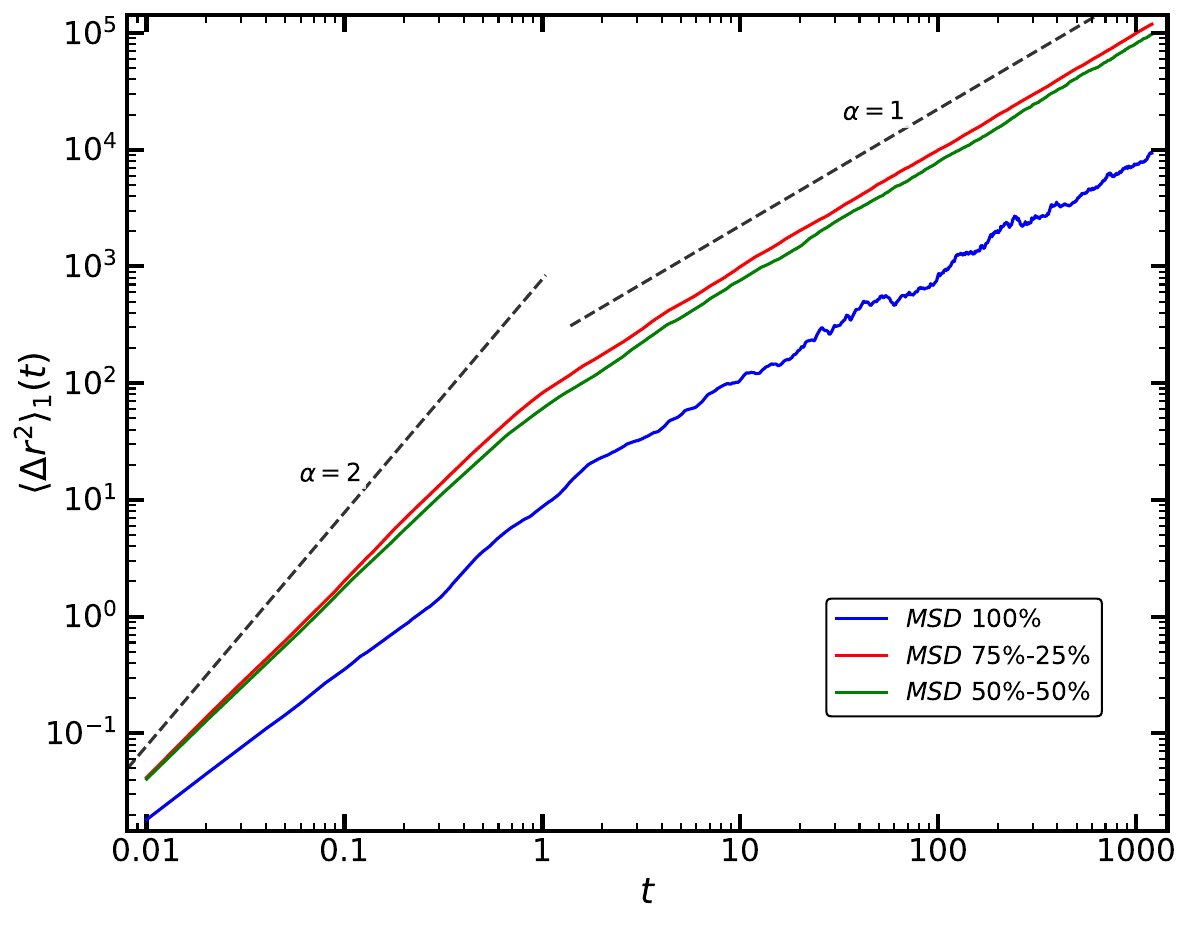}
   % \caption{Título 3}
    %\label{fig:msd3}
  \end{subfigure}\hfill
  \begin{subfigure}{0.48\textwidth}
    \centering
    \includegraphics[width=\linewidth]{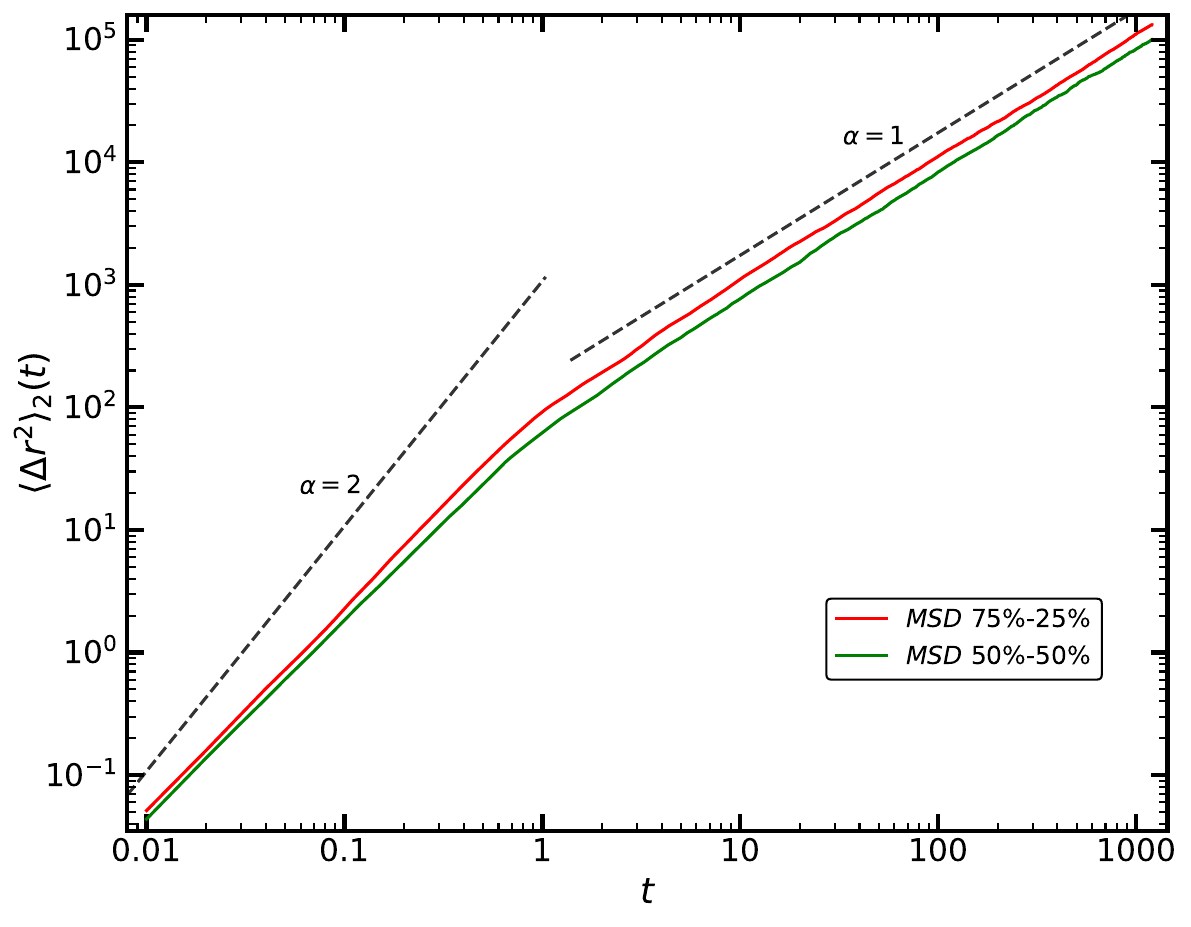}
  %  \caption{Título 4}
    %\label{fig:msd4}
  \end{subfigure}

  %\vspace{0.2ex}

  % ----- Fila 3 -----
  \begin{subfigure}{0.48\textwidth}
    \centering
    \includegraphics[width=\linewidth]{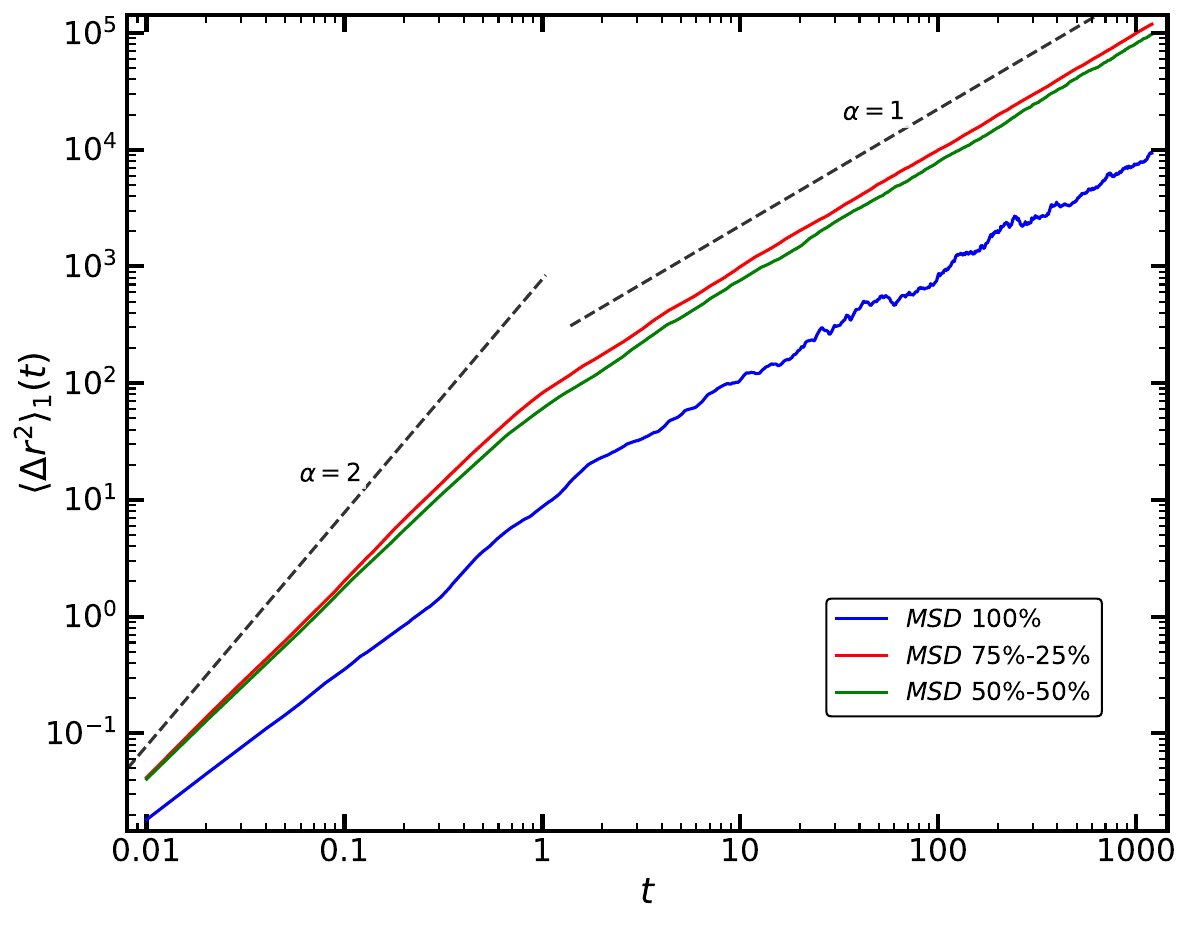}
 %   \caption{Título 5}
    %\label{fig:msd5}
  \end{subfigure}\hfill
  \begin{subfigure}{0.48\textwidth}
    \centering
    \includegraphics[width=\linewidth]{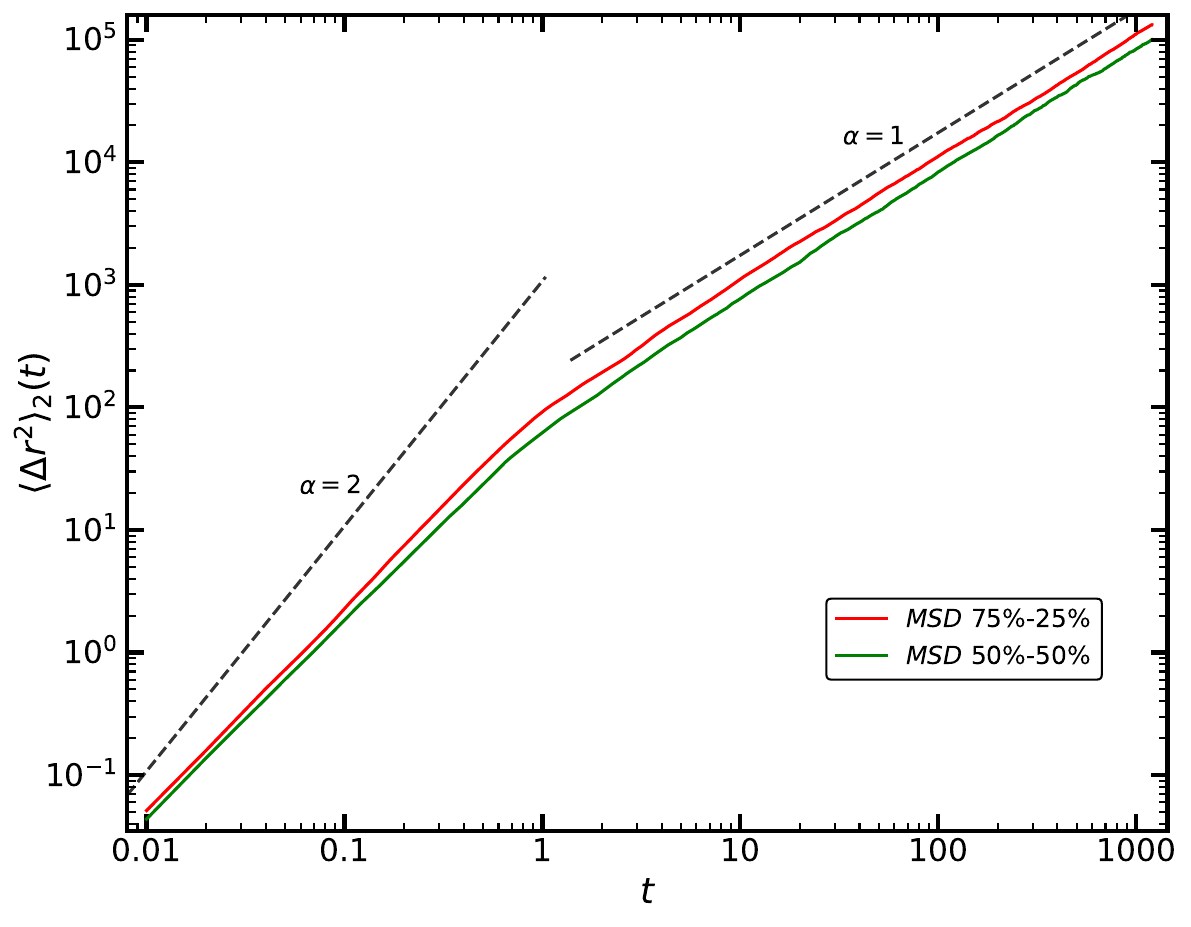}
   % \caption{Título 6}
    %\label{fig:msd6}
  \end{subfigure}

  \caption{Log-log plot of the mean-square displacements as a function of time in the dense coexisting state. In each plot, we represented the results for $X_1=1$ (blue lines), $0.75$ (red lines) and $0.25$ (green lines). The left column corresponds to the MSD of $1-$type particles and the right column to the MSD of $2-$type particles. From top to bottom: $Pe = 81, 100$ and $150$.}
  \label{fig:msd_2}
\end{figure*}

Now we turn to the MSD for the high-density coexisting states. Fig. \ref{fig:msd_2} shows their plots under the same conditions as those of the cases represented in Fig. \ref{fig:msd}. First, we note that, although in short-times the MSD seems to deviate slightly from the ballistic regime (which may be observed for shorter times), in all the cases it smoothly crossovers to a diffusive behavior for large times. Thus, there is no indication of caging effects associated with the emergence of jammed states. Now, the self-diffusion coefficients are significantly smaller than in the dilute state, as expected (see Table \ref{tab:DIFF}). On the other hand, it is now dependent on the mixture composition. In the binary mixtures, diffusion is higher for $X_1=0.75$ than for $X_1=0.5$ for each species and P\'eclet number. On the other hand, the monocomponent system shows diffusive behavior, with a much smaller self-diffusion coefficient than in the binary cases which decreases when the P\'eclet number increases. This is a surprising feature, since our analysis of the structural properties showed a spatial ordering which may indicate no large-$t$ diffusive behavior or caging effect. A possible solution of this puzzle is the presence of topological defects (mainly dislocations), which are highly mobile due to the particles activity, so diffusion of particles is driven by jump events when a topological defect is near a particle. This may be related to the phenomenology observed experimentally in active disks crystals, which are never dynamically arrested due to the formation of active droplets of active liquids which rapidly propagate across the system\cite{Briand2016} and that shows a macroscopic sheared flow under hexagonal confinement\cite{Briand2018}. This mechanism is much slower than the usual diffusion in binary mixtures, which corresponds to a more liquid-like dynamics. As there are fewer topological defects when $Pe$ increases (as it is indicated by the analysis of the hexatic order parameter, which is reduced as there are more topological defects), this could explain the dependence of the diffusion coefficients on the P\'eclet number in the monocomponent case. However, further work is needed to fully understand this phenomenon.

\section{\label{conclusions}Conclusions}

In this paper we present a preliminary study of the MIPS which occurs in a binary mixture of active soft Brownian particles by Brownian Dynamics. The two types of particles only differ on the interaction potential, sharing the same dynamical properties such as self-propelling speed and translational and rotational friction coefficients. We used Weeks-Chandler-Andersen interaction potentials between each pair of particles with parameters based on Lennard-Jones potentials in a glass-forming passive binary mixture. As a consequence, we observe that the high-density coexisting state is spatially disordered. This is in contrast with the monocomponent case, in which these dense states are solid-like. These findings are corroborated by the analysis of the radial distribution functions and the hexatic order parameters. Regarding the dynamical properties, the low- and high-density states show diffusive behavior at large times, indicating that they are fluid-like in the binary case. Surprisingly, diffusive behavior is also observed in the solid-like high-density coexisting states for the monocomponent system. This can be explained by the observation of active topological defects, mainly dislocations, which may drive particle diffusion. 

There are many open questions which will addressed in future works. The first question is how the high-density coexisting states become disordered as the $1-$type mole fraction decreases. In this respect, some preliminary results for a small number of $2-$type impurities in the otherwise $1-$type monocomponent system show that 5-fold disclinations nucleate in the impurities, breaking the hexatic ordering. Thus, we expect that, as $X_1$ decreases, the number of disclinations will increase until the hexatic ordering is completely lost, but how this occurs is unknown. A second question we would like to study is to characterize the activity of the topological defects in the solid-like high-density coexisting states for the monocomponent system and how this feature determines the particle diffusivity. 

\begin{acknowledgments}
This paper is dedicated to Carlos Vega on the occasion of his 60th birthday and his outstanding contributions in the field of Computational Statistical Physics. 
We wish to acknowledge financial support from the Agencia Estatal de Investigaci\'on (Ministerio de Ciencia, Innovaci\'on y Universidades, Spain) through 
Project No. PID2024-156257NB-C21. 
\end{acknowledgments}

\section*{Data Availability Statement}

The data that support the findings of this study are available from the corresponding author upon reasonable request.

\appendix
\section*{Appendix: Evaluation of Local Density Probability Distribution Functions.\label{appendix}}

In this paper we have characterized MIPS by the analysis of the density profiles calculated by simulations in elongated simulation cells which promote the formation of dense and dilute slabs for large P\'eclet numbers. However, in the literature MIPS has also been characterized by the local density probability distribution functions (PDFs). 
For this purpose, the simulation box is divided into a grid of rectangular bins of dimensions $5\times 4$ in reduced units. The local density for each bin is calculated as the ratio between the number of particles $N_i$ found in that bin by its area. We build a histogram for the densities obtained for each bin over configurations generated during the simulation. The local density PDF is obtained by normalizing appropriately this histogram. Under MIPS conditions, this distribution shows a bimodal shape with two maxima corresponding the densities of the coexisting states. The width of the maxima are found to be dependent on the dimensions of the bins: the larger the bin, the narrower the maxima.  

\begin{figure*}[t] % [t] o [b]; en dos columnas, figure* sólo va arriba/abajo
  \centering

  % ----- ROW 1 -----
  \begin{subfigure}{0.48\textwidth}
    \centering
    \includegraphics[width=\linewidth]{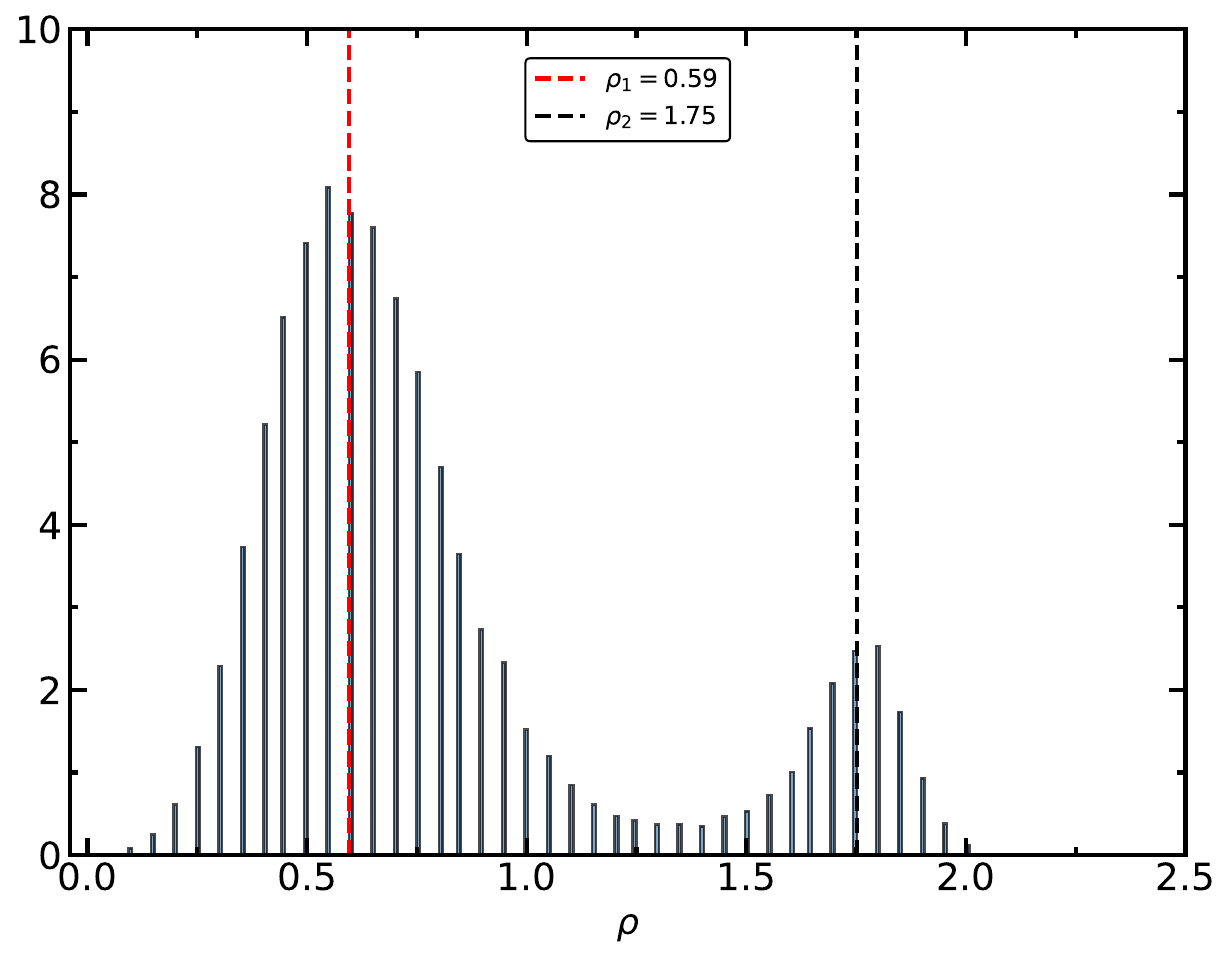}
    %\caption{}
    %\label{fig:PDF1}
  \end{subfigure}\hfill
  \begin{subfigure}{0.48\textwidth}
    \centering
    \includegraphics[width=\linewidth]{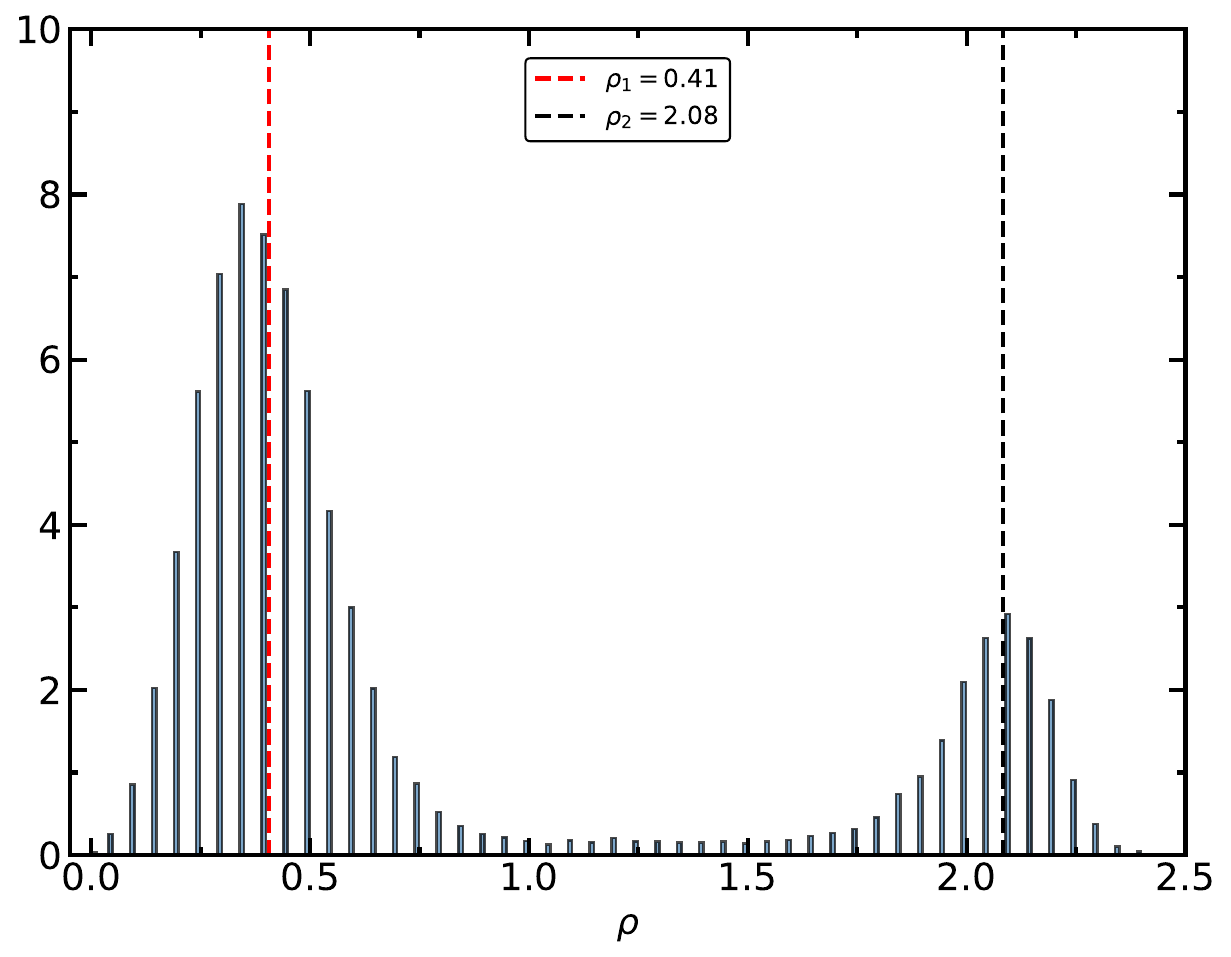}
    %\caption{}
    %\label{fig:PDF2}
  \end{subfigure}

  %\vspace{0.2ex}

  % ----- ROW 2 -----
  \begin{subfigure}{0.48\textwidth}
    \centering
    \includegraphics[width=\linewidth]{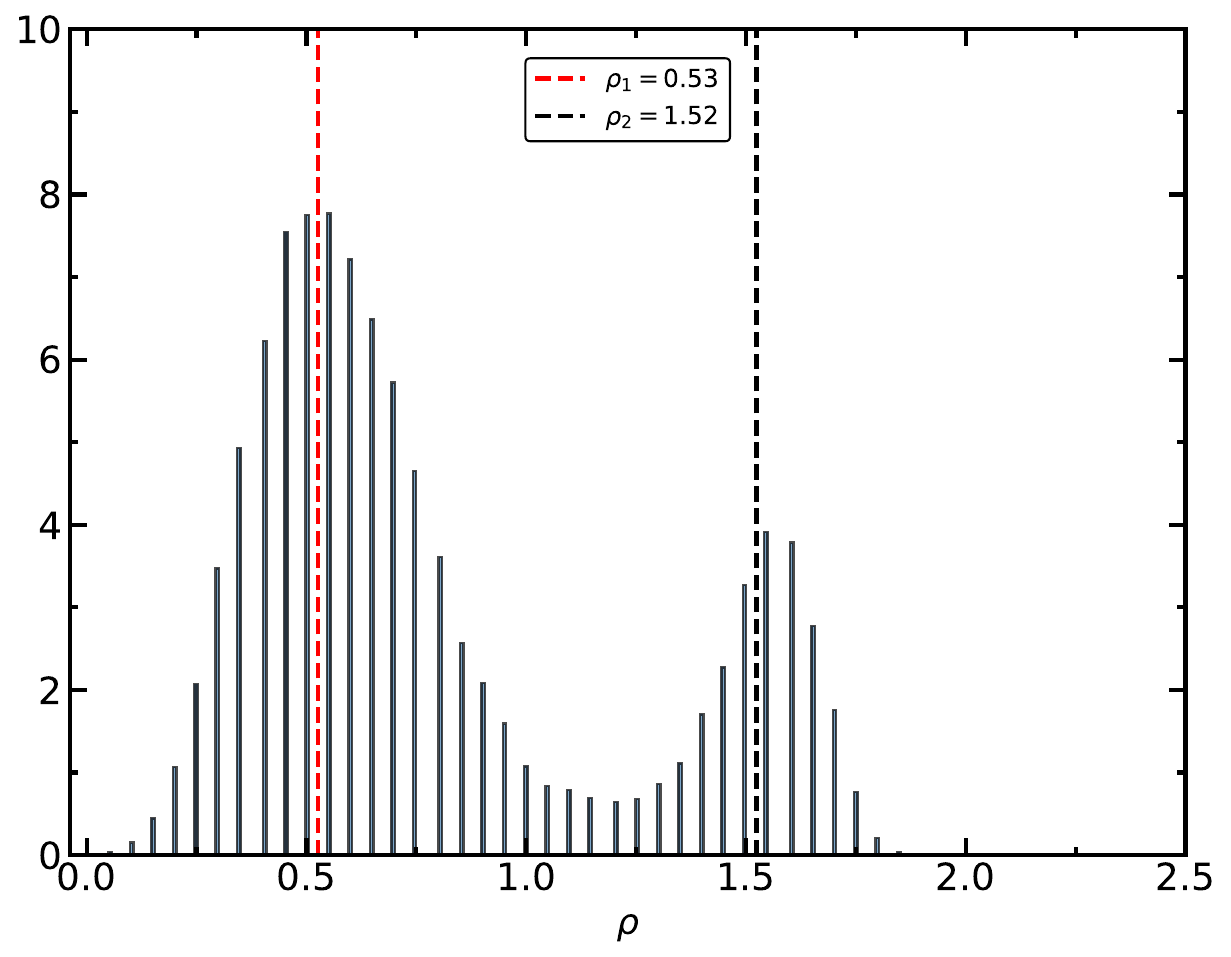}
   %\caption{}
    %\label{fig:PDF3}
  \end{subfigure}\hfill
  \begin{subfigure}{0.48\textwidth}
    \centering
    \includegraphics[width=\linewidth]{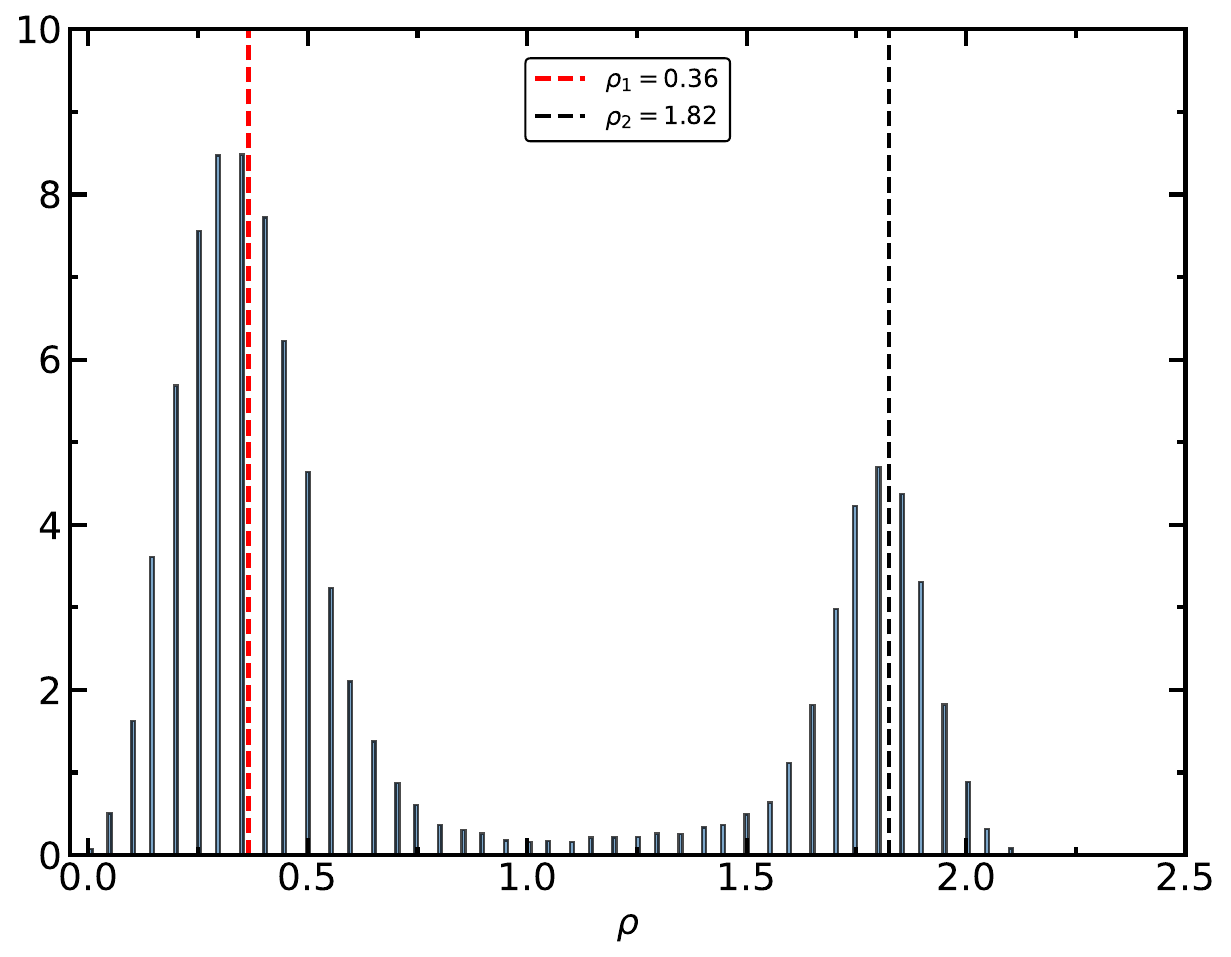}
  %  \caption{}
    %\label{fig:PDF4}
  \end{subfigure}

  %\vspace{0.2ex}

  % ----- ROW3 -----
  \begin{subfigure}{0.48\textwidth}
    \centering
    \includegraphics[width=\linewidth]{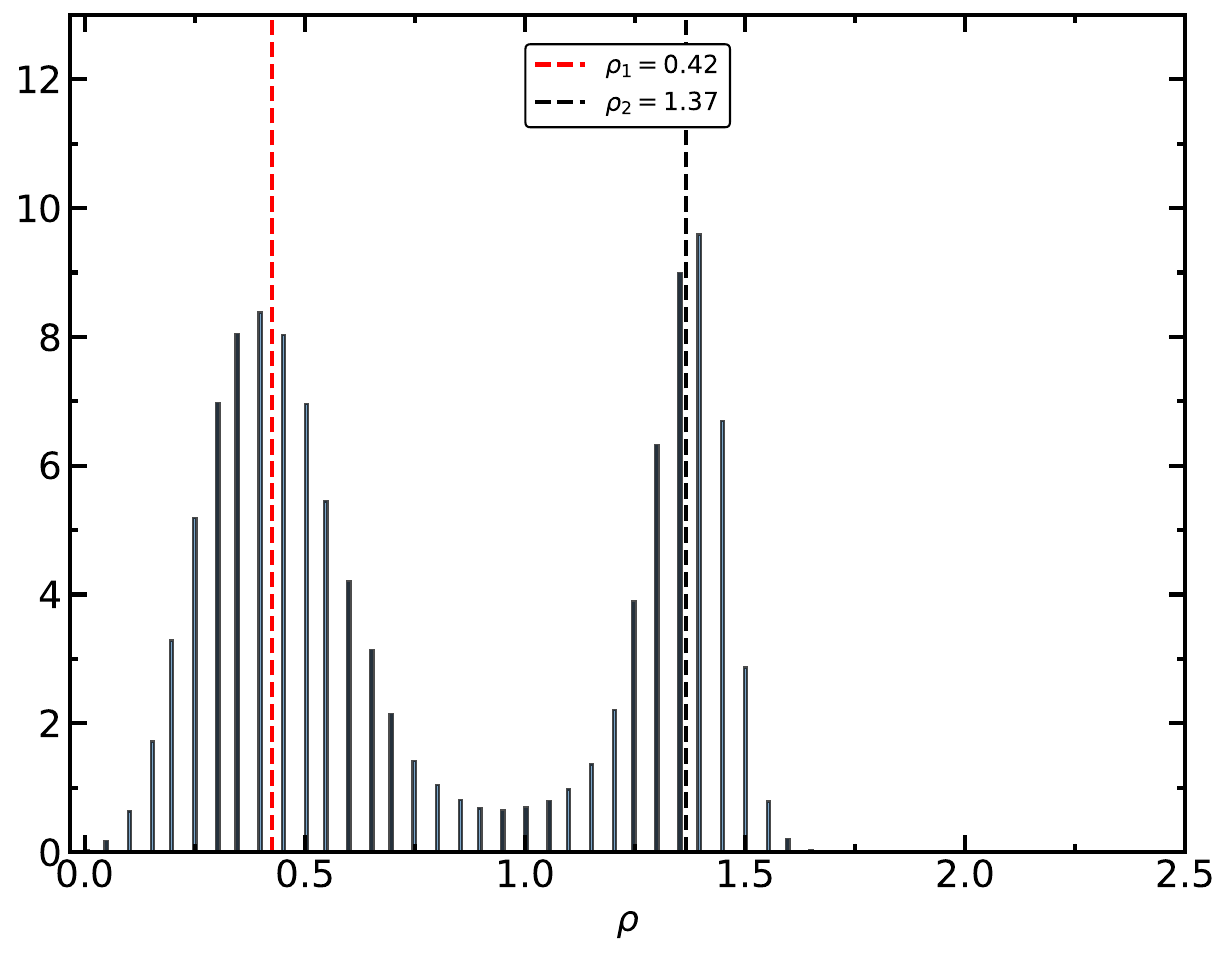}
 %   \caption{}
    %\label{fig:PDF5}
  \end{subfigure}\hfill
  \begin{subfigure}{0.48\textwidth}
    \centering
    \includegraphics[width=\linewidth]{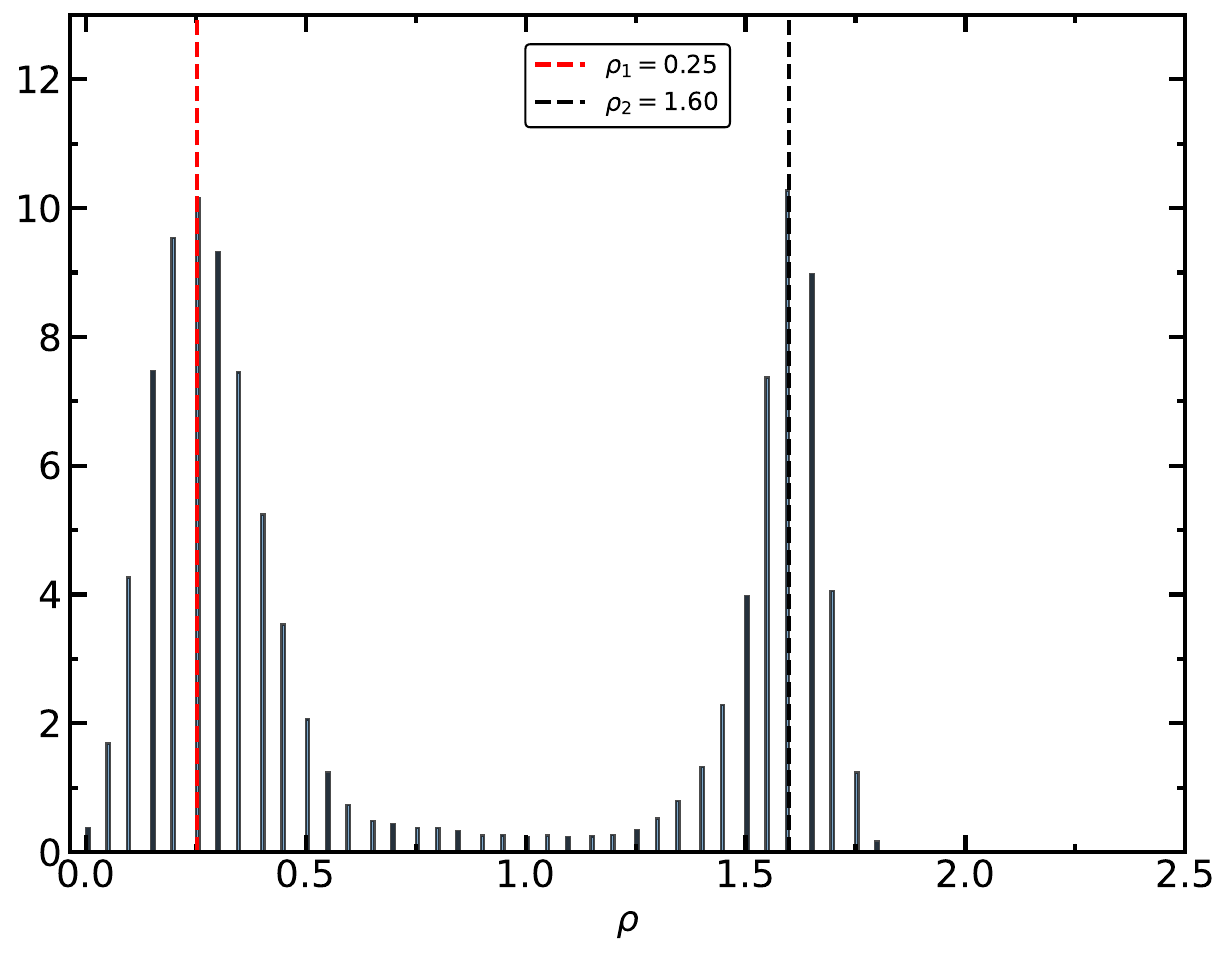}
   % \caption{}
    %\label{fig:PDF6}
  \end{subfigure}

   % \vspace{0.2ex}

  \caption{Local density probability distribution under MIPS conditions. From top to bottom, molar fractions $X_1=0.5$, $0.75$, and $1.00$ respectively. The left column corresponds to P\'eclet number $Pe=81$ and the right column corresponds to $Pe=150$. The vertical dashed lines correspond to the densities of the coexisting phases obtained from the density profile.}
  \label{fig:PDF}
\end{figure*}

Figure \ref{fig:PDF} plots the local density PDF for molar fractions $X_1=0.5$, $0.75$, and $1.00$ for values of the P\'eclet number $81$ and $150$. As we see, the positions of the maxima agree with the estimated densities obtained by averaging the density profiles along their plateaus, confirming that both methods are equivalent for $Pe>80$. 

\nocite{*}

\bibliography{biblio}% Produces the bibliography via BibTeX.

\end{document}